\documentclass[12pt,reqno]{article}
\usepackage{amsfonts}
\usepackage{amsmath}
\usepackage{amsbsy}
\usepackage{graphicx}
\usepackage{amssymb,latexsym}
 \numberwithin{equation}{section}

\begin{document}
 \allowdisplaybreaks[1]
\title{Decoupling Solution Moduli of Bigravity}
\author{Nejat Tevfik Y$\i$lmaz\\
Department of Electrical and Electronics Engineering,\\
Ya\c{s}ar University,\\
Sel\c{c}uk Ya\c{s}ar Kamp\"{u}s\"{u}\\
\"{U}niversite Caddesi, No:35-37,\\
A\u{g}a\c{c}l\i Yol, 35100,\\
Bornova, \.{I}zmir, Turkey.\\
\texttt{nejat.yilmaz@yasar.edu.tr}} \maketitle
\begin{abstract}
A complete classification of exact solutions of ghost-free,
massive bigravity is derived which enables the dynamical
decoupling of the background, and the foreground metrics. The
general decoupling solution space of the two metrics is
constructed. Within this branch of the solution space the
foreground metric theory becomes general relativity (GR) with an
additional effective cosmological constant, and the background
metric dynamics is governed by plain GR.
\end{abstract}

\section{Introduction}
Recently, a ghost-free \cite{BD1,BD2} nonlinear massive gravity
theory was constructed in \cite{dgrt1,dgrt2}. This theory is a
nonlinear generalization of \cite{fp}. Independently later on,
this ghost-free massive gravity with a flat reference metric was
also extended to include a general background metric in
\cite{hr1,hr2,hr3}. A ghost-free two-dynamical-metric theory,
namely the bigravity as a covering theory of the massive gravity
has also been proposed by introducing the dynamics for the
background metric \cite{hrbg,bac1,bac2,bac3}.

In this paper, by referring to the simple observation already
pointed out in \cite{bac1} which leads to a dynamical decoupling
of the background [$f$], and the foreground [$g$] metrics we will
derive the general solutions $f(x^\mu)=F(g(x^\mu),x^\mu)$ which
enable the two metrics to be solutions of two disjoint general
relativity (GR) theories. This is possible if a portion of the
effective energy-momentum tensor entering into the $g$-metric
equations as a course of the interaction Lagrangian between the
two metrics vanishes. As the same term also appears in the
$f$-metric equations by being the only contribution, when it
vanishes the sets of field equations of the two metrics completely
decouple from each other yielding only an algebraic matrix
equation which generates this picture. This matrix equation
written for $f$, and $g$ plays the role of a solution ansatz that
leads us to a branch of the solution space generated by a
Cartesian product of two GR's. This matrix ansatz equation will be
at the center of our analysis. In the following, we will derive
the general solutions of this cubic matrix equation when none of
the parameters of the theory vanish. Thus, we will be able to give
a complete description of the solution space $\Gamma[f,g]$ whose
elements lead to this dynamical decoupling of the two metric
sectors. We will also show that, the classification scheme of the
analytically available solutions $\{f,g\}$ admits a similarity
equivalence class structure.

In Section one, following a summary of the bigravity dynamics we
will obtain the decoupling ansatz matrix equation we have
mentioned above. Then, in the next section, we will derive the
general solutions of this cubic matrix equation for generic
constant coefficients. Since the coefficients in the actual
equation are functions of the elementary symmetric polynomials of
the solutions themselves a more refining analysis is needed.
Therefore, later on, we will present a parametric derivation which
enables us to construct not only the solutions of this involved
matrix equation, but also their elementary symmetric polynomials
as functions of the parameters of the theory. Subsequently, we
will show that a subset of the generic solutions for constant
coefficients must be omitted, when one insists on having an entire
set of nonzero theory parameters. Besides, some of the generic
solutions are forced to yield the same form when they are plugged
into the actual matrix equation we have. In Section four, we will
also present a formal definition of the decoupling solution space
$\Gamma[f,g]$ of the bigravity theory. We will show that, this
space contains a major subset that is composed of analytically
well-defined similarity equivalence classes of solutions. Finally,
in Section five we will explicitly construct the proportional
background solutions, and give an example of
Friedmann-Lemaitre-Robertson-Walker (FLRW) on FLRW case.
\section{The dynamical decoupling}
The action for the ghost-free bimetric gravity
\cite{hrbg,bac1,bac2,bac3} for the foreground, $g$, and the
background, $f$ metrics in the presence of two types of matter can
be given as
\begin{subequations}\label{e1}
\begin{align}
 S=&-\frac{1}{16\pi G}\int dx^4\sqrt{-g}\bigg[ R^g+\Lambda^g -2m^2\mathcal{L}_{int}(\sqrt{\Sigma})\bigg]+S_{M}^g\notag\\
 &-\frac{\kappa}{16\pi G}\int dx^4\sqrt{-f}\bigg[ R^f+\Lambda^f \bigg]+\epsilon
 S_{M}^f,\tag{\ref{e1}}
\end{align}
\end{subequations}
 where $R^g,R^f,\Lambda^g,\Lambda^f$ are the corresponding Ricci scalars, and the cosmological constants for the two metrics, respectively.
 $S_{M}^g,S_{M}^f$ are the two different types of matter which independently couple
to $g$, and $f$, respectively. The interaction Lagrangian of the
two metrics is
\begin{equation}\label{e2}
\mathcal{L}_{int}(\sqrt{\Sigma})=\beta_{1}
e_{1}(\sqrt{\Sigma})+\beta_{2} e_{2}(\sqrt{\Sigma})+\beta_{3}
e_{3}(\sqrt{\Sigma}),
\end{equation}
where $\{e_n\}$ are the elementary symmetric polynomials
\begin{subequations}\label{e3}
\begin{align}
e_{1}\equiv e_{1}(\sqrt{\Sigma})&=tr\sqrt{\Sigma},\notag\\
e_{2}\equiv e_{2}(\sqrt{\Sigma})&=\frac{1}{2}\big((tr\sqrt{\Sigma})^2-tr(\sqrt{\Sigma})^2\big),\notag\\
e_{3}\equiv
e_{3}(\sqrt{\Sigma})&=\frac{1}{6}\big((tr\sqrt{\Sigma})^3-3\,tr\sqrt{\Sigma}\:tr(\sqrt{\Sigma})^2+2\,tr(\sqrt{\Sigma})^3\big),
\tag{\ref{e3}}
\end{align}
\end{subequations}
of the square-root-matrix
\begin{equation}\label{e3.1}
\sqrt{\Sigma}=\sqrt{g^{-1}f}.
\end{equation}
Likewise in \cite{bac3} the original interaction terms $\beta_0
e_0=\beta_0$, and $\beta_4 e_4=\beta_4\text{det}\sqrt{\Sigma}$ are
trivially plugged into the cosmological constants $\Lambda_g$, and
$\Lambda_f$, respectively. If we demand that Eq.\eqref{e2} gives
the Fierz-Pauli form in the weak-field limit then we must have
\cite{hrbg}
\begin{equation}\label{e3.2}
\beta_1+2\beta_2+\beta_3=-1.
\end{equation}
Varying Eq.\eqref{e1} with respect to $g$ gives the $g$-equation
\begin{equation}\label{e4}
R^g_{\mu\nu}-\frac{1}{2}R^g g_{\mu\nu}-\frac{1}{2}\Lambda^g
g_{\mu\nu}-m^2\mathcal{T}^g_{\mu\nu}=8\pi GT^{g}_{M\,\mu\nu}.
\end{equation}
Also, variation with respect to $f$ results in the $f$-equation
\begin{equation}\label{e4.1}
\kappa\big[R^f_{\mu\nu}-\frac{1}{2}R^f
f_{\mu\nu}-\frac{1}{2}\Lambda^f
f_{\mu\nu}\big]-m^2\mathcal{T}^f_{\mu\nu}=\epsilon 8\pi
GT^{f}_{M\,\mu\nu}.
\end{equation}
In these field equations the contributions coming from the
interaction term that is given in Eq.\eqref{e2} are the effective
energy-momentum tensors
\begin{equation}\label{e5}
\mathcal{T}^g_{\mu\nu}=g_{\mu\rho}\tau^{\rho}_{\:\:\nu}-\mathcal{L}_{int}g_{\mu\nu},
\end{equation}
and
\begin{equation}\label{e6}
\mathcal{T}^f_{\mu\nu}=-\frac{\sqrt{-g}}{\sqrt{-f}}f_{\mu\rho}\tau^{\rho}_{\:\:\nu},
\end{equation}
respectively. Here $\{\tau^{\rho}_{\:\:\nu}\}$ are the elements of
the matrix $\tau$ \cite{bac3}
\begin{equation}\label{e7}
\tau=\beta_3(\sqrt{\Sigma})^3-(\beta_2+\beta_3
e_1)(\sqrt{\Sigma})^2+(\beta_1+\beta_2 e_1+\beta_3
e_2)\sqrt{\Sigma},
\end{equation}
namely $\tau^\rho_{\:\:\nu}\equiv [\tau]^\rho_{\:\:\nu}$. Both of
the effective energy-momentum tensors must be covariantly constant
\begin{equation}\label{e7.6}
\nabla^g_\mu(\mathcal{T}^{g})^{\mu}_{\:\:\nu}=0,\quad
\nabla^f_\mu(\mathcal{T}^{f})^{\mu}_{\:\:\nu}=0.
\end{equation}
If one of these constraints is satisfied then the other one is
automatically satisfied \cite{bm2,bm1}. As discussed in
\cite{bac1} if one chooses
\begin{equation}\label{e8}
\tau=0,
\end{equation}
then dynamically the Eqs.\eqref{e4}, and \eqref{e4.1} decouple
from each other. In this case the first of the constraints
\eqref{e7.6} gives
\begin{equation}\label{e8.1}
\partial_\mu\mathcal{L}_{int}=0,
\end{equation}
and thus, as the solution we will take
\begin{equation}\label{e8.2}
\mathcal{L}_{int}=-\frac{1}{2}\tilde{\Lambda}.
\end{equation}
Therefore, from Eqs.\eqref{e5}, \eqref{e6} we have
\begin{equation}\label{e8.3}
\mathcal{T}^g_{\mu\nu}=\frac{1}{2}\tilde{\Lambda}g_{\mu\nu},\quad
\mathcal{T}^f_{\mu\nu}=0.
\end{equation}
Consequently, the $g$-equation Eq.\eqref{e4} becomes the usual
Einstein equations for $g$ with an additional effective
cosmological constant $\tilde{\Lambda}$, whereas the
dynamically-disjoint $f$-equation Eq.\eqref{e4.1} reduces to be
coupling-constant-modified Einstein equations for $f$. The rest of
our analysis will be devoted to find the general solutions of the
matrix equation\footnote{We will prefer to work with the negative
of the Eq.\eqref{e8} in accordance with the massive gravity
formalism \cite{massgrav,mgrphysfluid}, and for future relevance.}
\begin{equation}\label{e10}
\sqrt{\Sigma}\bigg(A(\sqrt{\Sigma})^2+B(\sqrt{\Sigma})+C\mathbf{1}_4\bigg)=0,
\end{equation}
where
\begin{subequations}\label{e11}
\begin{align}
A&=-\beta_3,\notag\\
B&=\beta_2+\beta_3e_1,\notag\\
C&=-\beta_1-\beta_2e_1-\beta_3e_2,\tag{\ref{e11}}
\end{align}
\end{subequations}
which constitute the effective solution space $\Gamma[g,f]$ of the
ghost-free bigravity action \eqref{e1} that enables the
above-mentioned dynamical decoupling for the foreground, and the
background metrics.
\section{The structure of the solution space $\Gamma$}
Now, let us consider the matrix equation
\begin{equation}\label{e12}
AX^3+BX^2+CX=X(AX^2+BX+C\mathbf{1}_{4})=X(X-\lambda_1\mathbf{1}_{4})(X-\lambda_2\mathbf{1}_{4})=0,
\end{equation}
for a $4\times 4$ matrix function $X(x^\mu)$. For the following
analysis we will disregard the solutions which require either of
the $\beta$-coefficients to be zero. The characteristic polynomial
of any $4\times 4$ matrix $X$ would be the degree-four polynomial
\begin{equation}\label{e13}
P_X(t)=det(t\mathbf{1}_{4}-X),
\end{equation}
whose four roots are the eigenvalues of $X$. We should observe
first that if $X$ is a solution of Eq.\eqref{e12} then for any
invertible matrix function $P(x^\mu)$,  $P^{-1}XP$ is also a
solution. Therefore, in order to find the general solutions of
Eq.\eqref{e12} it would be sufficient to classify the Jordan
canonical forms satisfying \eqref{e12}. Our main objective next,
will be the classification of the similarity equivalence classes
of solutions with respect to their minimum polynomials $m(X)$. The
roots of $m(X)$ are the same with the eigenvalues of the various
equivalence classes of matrices having that minimum polynomial
with differing multiplicities of course. Since Eq.\eqref{e12} is a
degree-three polynomial equation when it is factorized, its
various factors with degrees smaller than or equal to three will
define the minimum polynomials of its $4\times 4$ matrix function
solutions. In other words, the solutions can be classified with
respect to these minimum polynomial factors.

\subsection{The algebraic structure}
In the following classification, we will identify the entire set
of similarity equivalence classes of solutions satisfying
Eq.\eqref{e12} by simply taking the coefficients in Eq.\eqref{e12}
to be constants. We will group the solutions with respect to their
minimum polynomials and it will be sufficient to determine the
Jordan canonical form spectrum of each minimum polynomial which
corresponds to some combination of the factors in Eq.\eqref{e12}.
\begin{enumerate}
\item{\textsl{Solutions with $m(X)=X$}}

Since now, the minimum polynomial has no repeated roots and its
unique root is zero these solutions are diagonalizable with zero
eigenvalues. They are the trivial solutions $X=0$. These solutions
would demand at least one of the metrics to be zero via
Eq\eqref{e3.1} as metrics are invertible matrices.

Next, we will classify the solutions of Eq.\eqref{e12} with
respect to the root structure of the quadratic factor.
\begin{itemize}
\item\title{\textbf{\underline{The cases when $B^2-4AC>0$\:\: :}}}
\end{itemize}
In this case, the factor $AX^2+BX+C\mathbf{1}_{4}$ has two
distinct real roots $\lambda_1,\lambda_2$.

\item{\textsl{Solutions with
$m(X)=X-\lambda_{1,2}$$\mathbf{1}_{4}$}}

 Now, there exist diagonal
solutions as the minimum polynomials have no repeated roots and
they are linear. These solutions are
\begin{equation}\label{e14}
U_1=\lambda_1\mathbf{1}_{4},\quad\quad
U_2=\lambda_2\mathbf{1}_{4},
\end{equation}
with minimum polynomials $m(X)=X-\lambda_1\mathbf{1}_{4}$,
$m(X)=X-\lambda_2\mathbf{1}_{4}$, respectively.

\item{\textsl{Solutions with $m(X)=AX^2+BX+C\mathbf{1}_{4}$}}

The matrices with a minimum polynomial of the form
$m(X)=AX^2+BX+C\mathbf{1}_{4}$ are diagonalizable when the roots
are distinct like our case here. Thus, by considering all the
possible multiplicities of the eigenvalues which are the same with
these distinct roots $\lambda_1,\lambda_2$ the Jordan forms
corresponding to this minimum polynomial become
\begin{subequations}\label{e15}
\begin{align}
 U_3=\left(\begin{matrix}
\lambda_{1}&\texttt{0}&\texttt{0}&\texttt{0}
\\
\texttt{0}&\lambda_2&\texttt{0}&\texttt{0}\\\texttt{0}&\texttt{0}&\lambda_1
&\texttt{0}\\\texttt{0}&\texttt{0}
&\texttt{0}&\lambda_2\end{matrix}\right), U_4=\left(\begin{matrix}
\lambda_1&\texttt{0}&\texttt{0}&\texttt{0}
\\
\texttt{0}&\lambda_1&\texttt{0}&\texttt{0}\\\texttt{0}&\texttt{0}&\lambda_1
&\texttt{0}\\\texttt{0}&\texttt{0}
&\texttt{0}&\lambda_2\end{matrix}\right),\notag\\
U_5= \left(\begin{matrix}
\lambda_2&\texttt{0}&\texttt{0}&\texttt{0}
\\
\texttt{0}&\lambda_2&\texttt{0}&\texttt{0}\\\texttt{0}&\texttt{0}&\lambda_2
&\texttt{0}\\\texttt{0}&\texttt{0}
&\texttt{0}&\lambda_1\end{matrix}\right).\quad\quad\quad\quad\quad\quad\tag{\ref{e15}}
\end{align}
\end{subequations}
\item{\textsl{Solutions with
$m(X)=X(X-\lambda_{1}$$\mathbf{1}_{4})$}}

In this case, the roots of the minimum polynomial are
$\{0,\lambda_1\}$. Again, the matrices with this minimum
polynomial are diagonalizable. The possible Jordan forms with
$\{0,\lambda_1\}$ eigenvalues read
\begin{equation}\label{e16}
 U_6=\left(\begin{matrix}
\lambda_{1}&\texttt{0}&\texttt{0}&\texttt{0}
\\
\texttt{0}&\texttt{0}&\texttt{0}&\texttt{0}\\\texttt{0}&\texttt{0}&\texttt{0}
&\texttt{0}\\\texttt{0}&\texttt{0}
&\texttt{0}&\texttt{0}\end{matrix}\right),
U_7=\left(\begin{matrix}
\lambda_1&\texttt{0}&\texttt{0}&\texttt{0}
\\
\texttt{0}&\lambda_1&\texttt{0}&\texttt{0}\\\texttt{0}&\texttt{0}&\texttt{0}
&\texttt{0}\\\texttt{0}&\texttt{0}
&\texttt{0}&\texttt{0}\end{matrix}\right), U_8=
\left(\begin{matrix} \lambda_1&\texttt{0}&\texttt{0}&\texttt{0}
\\
\texttt{0}&\lambda_1&\texttt{0}&\texttt{0}\\\texttt{0}&\texttt{0}&\lambda_1
&\texttt{0}\\\texttt{0}&\texttt{0}
&\texttt{0}&\texttt{0}\end{matrix}\right).
\end{equation}
\item{\textsl{Solutions with
$m(X)=X(X-\lambda_{2}$$\mathbf{1}_{4})$}}

Similarly, the diagonalizable solutions with eigenvalues
$\{0,\lambda_2\}$ as the roots of the minimum polynomial have the
following possible Jordan forms
\begin{equation}\label{e17}
 U_9=\left(\begin{matrix}
\lambda_{2}&\texttt{0}&\texttt{0}&\texttt{0}
\\
\texttt{0}&\texttt{0}&\texttt{0}&\texttt{0}\\\texttt{0}&\texttt{0}&\texttt{0}
&\texttt{0}\\\texttt{0}&\texttt{0}
&\texttt{0}&\texttt{0}\end{matrix}\right),
U_{10}=\left(\begin{matrix}
\lambda_2&\texttt{0}&\texttt{0}&\texttt{0}
\\
\texttt{0}&\lambda_2&\texttt{0}&\texttt{0}\\\texttt{0}&\texttt{0}&\texttt{0}
&\texttt{0}\\\texttt{0}&\texttt{0}
&\texttt{0}&\texttt{0}\end{matrix}\right), U_{11}=
\left(\begin{matrix} \lambda_2&\texttt{0}&\texttt{0}&\texttt{0}
\\
\texttt{0}&\lambda_2&\texttt{0}&\texttt{0}\\\texttt{0}&\texttt{0}&\lambda_2
&\texttt{0}\\\texttt{0}&\texttt{0}
&\texttt{0}&\texttt{0}\end{matrix}\right).
\end{equation}

\item{\textsl{Solutions with
$m(X)=X(X-\lambda_{1}$$\mathbf{1}_{4})(X-\lambda_{2}$$\mathbf{1}_{4})$}}

Now, the roots of the minimum polynomial are
$\{0,\lambda_1,\lambda_2\}$. They are distinct, and this states
that the solutions whose minimum polynomial is of this form must
be again diagonalizable. The possible Jordan forms constructed
from these eigenvalues are
\begin{subequations}\label{e18}
\begin{align}
 U_{12}=\left(\begin{matrix}
\texttt{0}&\texttt{0}&\texttt{0}&\texttt{0}
\\
\texttt{0}&\lambda_1&\texttt{0}&\texttt{0}\\\texttt{0}&\texttt{0}&\lambda_1
&\texttt{0}\\\texttt{0}&\texttt{0}
&\texttt{0}&\lambda_2\end{matrix}\right),
U_{13}=\left(\begin{matrix}
\texttt{0}&\texttt{0}&\texttt{0}&\texttt{0}
\\
\texttt{0}&\texttt{0}&\texttt{0}&\texttt{0}\\\texttt{0}&\texttt{0}&\lambda_1
&\texttt{0}\\\texttt{0}&\texttt{0}
&\texttt{0}&\lambda_2\end{matrix}\right),\notag\\ U_{14}=
\left(\begin{matrix} \texttt{0}&\texttt{0}&\texttt{0}&\texttt{0}
\\
\texttt{0}&\lambda_2&\texttt{0}&\texttt{0}\\\texttt{0}&\texttt{0}&\lambda_2
&\texttt{0}\\\texttt{0}&\texttt{0}
&\texttt{0}&\lambda_1\end{matrix}\right).\quad\quad\quad\quad\quad\quad\tag{\ref{e18}}
\end{align}
\end{subequations}
\begin{itemize}
\item\title{\textbf{\underline{The cases when $B^2-4AC=0$\:\: :}}}
\end{itemize}
For these cases, the factor $AX^2+BX+C\mathbf{1}_{4}$ has a
repeated real root which we will call $\lambda^{\prime}$.

\item{\textsl{Solutions with
$m(X)=X-\lambda^{\prime}$$\mathbf{1}_{4}$}}

 This solution, is another diagonal
one due to its minimum polynomial. It is
\begin{equation}\label{e18.5}
V_0=\lambda^{\prime}\mathbf{1}_{4}.
\end{equation}

\item{\textsl{Solutions with
$m(X)=(X-\lambda^{\prime}$$\mathbf{1}_{4})^2$}}

Since now, the minimum polynomial has repeated roots the matrices
with such a minimum polynomial are nondiagonalizable. Their
eigenvalues must be $\lambda^{\prime}$ with multiplicity four.
Thus the primary block is four dimensional and since the
multiplicity of the roots of $m(X)$ is two the maximum dimension
of the secondary block must be two. The possible Jordan forms are

\begin{subequations}\label{e19}
\begin{align}
 V_{1}&=\left(\begin{matrix}
\lambda^{\prime}&\texttt{1}&\texttt{0}&\texttt{0}
\\
\texttt{0}&\lambda^{\prime}&\texttt{0}&\texttt{0}\\\texttt{0}&\texttt{0}&\lambda^{\prime}
&\texttt{0}\\\texttt{0}&\texttt{0}
&\texttt{0}&\lambda^{\prime}\end{matrix}\right),\quad
V_{2}=\left(\begin{matrix}
\lambda^{\prime}&\texttt{0}&\texttt{0}&\texttt{0}
\\
\texttt{0}&\lambda^{\prime}&\texttt{1}&\texttt{0}\\\texttt{0}&\texttt{0}&\lambda^{\prime}
&\texttt{0}\\\texttt{0}&\texttt{0}
&\texttt{0}&\lambda^{\prime}\end{matrix}\right),\notag\\
V_{3}&= \left(\begin{matrix}
\lambda^{\prime}&\texttt{0}&\texttt{0}&\texttt{0}
\\
\texttt{0}&\lambda^{\prime}&\texttt{0}&\texttt{0}\\\texttt{0}&\texttt{0}&\lambda^{\prime}
&\texttt{1}\\\texttt{0}&\texttt{0}
&\texttt{0}&\lambda^{\prime}\end{matrix}\right),\quad
V_4=\left(\begin{matrix}
\lambda^{\prime}&\texttt{1}&\texttt{0}&\texttt{0}
\\
\texttt{0}&\lambda^{\prime}&\texttt{0}&\texttt{0}\\\texttt{0}&\texttt{0}&\lambda^{\prime}
&\texttt{1}\\\texttt{0}&\texttt{0}
&\texttt{0}&\lambda^{\prime}\end{matrix}\right).\tag{\ref{e19}}
\end{align}
\end{subequations}
We should remark that, it would be enough to take either of
$\{V_1,V_2,V_3\}$ however, to be on the safe side we include all
of them.

\item{\textsl{Solutions with
$m(X)=X(X-\lambda^{\prime}$$\mathbf{1}_{4})^2$}}

The roots of the minimum polynomial in this class of solutions are
$\{0,\lambda^{\prime}, \lambda^{\prime}\}$. Once more, $m(X)$ has
repeated roots so the matrices with this minimum polynomial are
nondiagonalizable. The dimension of the secondary blocks for the
eigenvalues $\lambda^{\prime}$, and $0$ are two, and one,
respectively. The possible Jordan canonical forms are
\begin{subequations}\label{e20}
\begin{align}
 V_{5}=\left(\begin{matrix}
\lambda^{\prime}&\texttt{1}&\texttt{0}&\texttt{0}
\\
\texttt{0}&\lambda^{\prime}&\texttt{0}&\texttt{0}\\\texttt{0}&\texttt{0}&\lambda^{\prime}
&\texttt{0}\\\texttt{0}&\texttt{0}
&\texttt{0}&\texttt{0}\end{matrix}\right),
V_{6}=\left(\begin{matrix}
\lambda^{\prime}&\texttt{0}&\texttt{0}&\texttt{0}
\\
\texttt{0}&\lambda^{\prime}&\texttt{1}&\texttt{0}\\\texttt{0}&\texttt{0}&\lambda^{\prime}
&\texttt{0}\\\texttt{0}&\texttt{0}
&\texttt{0}&\texttt{0}\end{matrix}\right),\notag\\
V_{7}= \left(\begin{matrix}
\lambda^{\prime}&\texttt{1}&\texttt{0}&\texttt{0}
\\
\texttt{0}&\lambda^{\prime}&\texttt{0}&\texttt{0}\\\texttt{0}&\texttt{0}&\texttt{0}
&\texttt{0}\\\texttt{0}&\texttt{0}
&\texttt{0}&\texttt{0}\end{matrix}\right).\quad\quad\quad\quad\quad\tag{\ref{e20}}
\end{align}
\end{subequations}
Like the previous case, we include both $V_5$, and $V_6$ in spite
of the fact that one would be enough to generate the similarity
class of solutions.

 \item{\textsl{Solutions with
$m(X)=X(X-\lambda^{\prime}$$\mathbf{1}_{4})$}}

There are no repeated roots of $m(X)$ for these solutions.
Therefore, the matrices with this minimum polynomial are
diagonalizable. The possible Jordan normal forms with the
eigenvalues $\{0,\lambda^{\prime}\}$ are
\begin{subequations}\label{e21}
\begin{align}
 V_{8}=\left(\begin{matrix}
\lambda^{\prime}&\texttt{0}&\texttt{0}&\texttt{0}
\\
\texttt{0}&\texttt{0}&\texttt{0}&\texttt{0}\\\texttt{0}&\texttt{0}&\texttt{0}
&\texttt{0}\\\texttt{0}&\texttt{0}
&\texttt{0}&\texttt{0}\end{matrix}\right),
V_{9}=\left(\begin{matrix}
\lambda^{\prime}&\texttt{0}&\texttt{0}&\texttt{0}
\\
\texttt{0}&\lambda^{\prime}&\texttt{0}&\texttt{0}\\\texttt{0}&\texttt{0}&\texttt{0}
&\texttt{0}\\\texttt{0}&\texttt{0}
&\texttt{0}&\texttt{0}\end{matrix}\right),\notag\\
V_{10}= \left(\begin{matrix}
\lambda^{\prime}&\texttt{0}&\texttt{0}&\texttt{0}
\\
\texttt{0}&\lambda^{\prime}&\texttt{0}&\texttt{0}\\\texttt{0}&\texttt{0}&\lambda^{\prime}
&\texttt{0}\\\texttt{0}&\texttt{0}
&\texttt{0}&\texttt{0}\end{matrix}\right).\quad\quad\quad\quad\quad\tag{\ref{e21}}
\end{align}
\end{subequations}
\begin{itemize}
\item\title{\textbf{\underline{The cases when $B^2-4AC<0$\:\: :}}}
\end{itemize}
For this class of solutions, the factor $AX^2+BX+C\mathbf{1}_{4}$
has complex roots which we will call $\lambda,\lambda^{*}$. In
this respect for such cases, the minimum polynomial of the
similarity equivalence classes of real solutions must include the
quadratic factor as a whole since for real matrices if $\lambda$
is an eigenvalue $\lambda^{*}$ must also be an eigenvalue, and
vice versa. For this reason, there are two possible minimum
polynomials and the corresponding solutions are nondiagonalizable.

\item{\textsl{Solutions with
$m(X)=(X-\lambda$$\mathbf{1}_{4})(X-\lambda^{*}\mathbf{1}_{4})$}}

The eigenvalues of the matrices with this $m(X)$ are
$\{\lambda,\lambda^{*}\}$. If we define the real and the imaginary
parts of $\lambda=R+I\mathbf{i}$, then the unique Jordan canonical
form for the real solutions with this minimum polynomial becomes
\begin{equation}\label{e22}
 Y=\left(\begin{matrix}
R&I&\texttt{0}&\texttt{0}
\\
-I&R&\texttt{0}&\texttt{0}\\\texttt{0}&\texttt{0}&R
&I\\\texttt{0}&\texttt{0} &-I&R\end{matrix}\right).
\end{equation}

\item{\textsl{Solutions with
$m(X)=X(X-\lambda$$\mathbf{1}_{4})(X-\lambda^{*}\mathbf{1}_{4})$}}

The eigenvalues of the solutions of this equivalence class are
$\{0,\lambda,\lambda^{*}\}$. The only possible Jordan normal form
for the real matrices satisfying
$m(X)=X(X-\lambda$$\mathbf{1}_{4})(X-\lambda^{*}\mathbf{1}_{4})=0$
is of the form

\begin{equation}\label{e23}
 Y^\prime =\left(\begin{matrix}
\texttt{0}&\texttt{0}&\texttt{0}&\texttt{0}
\\
\texttt{0}&\texttt{0}&\texttt{0}&\texttt{0}\\\texttt{0}&\texttt{0}&R
&I\\\texttt{0}&\texttt{0} &-I&R\end{matrix}\right).
\end{equation}
For the purpose of achieving the most general construction of the
solution space we should also consider the cases when the
coefficients in Eq.\eqref{e12} vanish. We will disregard the cases
when $A=0$ as they correspond to choosing $\beta_3=0$.

\begin{itemize}
\item\title{\textbf{\underline{The cases when $B=0$\:\: :}}}
\end{itemize}
If we assume that $C/A<0\:$\footnote{We will comment on the case
when $C/A>0$ in the next subsection.} then the quadratic factor in
Eq.\eqref{e12} can be written as
$(X-\tilde{\lambda})(X+\tilde{\lambda})$ where
$\tilde{\lambda}=\sqrt{-C/A}$.

\item{\textsl{Solutions with
$m(X)=X\pm\tilde{\lambda}$$\mathbf{1}_{4}$}}

There exist diagonal solutions. They are
\begin{equation}\label{e23.5}
K_1=\tilde{\lambda}\mathbf{1}_{4},\quad\quad
K_2=-\tilde{\lambda}\mathbf{1}_{4},
\end{equation}
with the minimum polynomials
$m(X)=X-\tilde{\lambda}\mathbf{1}_{4}$, and
$m(X)=X+\tilde{\lambda}\mathbf{1}_{4}$, respectively.

\item{\textsl{Solutions with $m(X)=AX^2+C\mathbf{1}_{4}$}}

The roots of the  minimum polynomial thus, the eigenvalues in this
case are $\{\tilde{\lambda},-\tilde{\lambda}\}$. They are distinct
hence, the matrices with this minimum polynomial are
diagonalizable. The possible Jordan forms are
\begin{subequations}\label{e24}
\begin{align}
 K_3=\left(\begin{matrix}
\tilde{\lambda}&\texttt{0}&\texttt{0}&\texttt{0}
\\
\texttt{0}&\tilde{\lambda}&\texttt{0}&\texttt{0}\\\texttt{0}&\texttt{0}&\tilde{\lambda}
&\texttt{0}\\\texttt{0}&\texttt{0}
&\texttt{0}&-\tilde{\lambda}\end{matrix}\right),
K_{4}=\left(\begin{matrix}
\tilde{\lambda}&\texttt{0}&\texttt{0}&\texttt{0}
\\
\texttt{0}&\tilde{\lambda}&\texttt{0}&\texttt{0}\\\texttt{0}&\texttt{0}&-\tilde{\lambda}
&\texttt{0}\\\texttt{0}&\texttt{0}
&\texttt{0}&-\tilde{\lambda}\end{matrix}\right),\notag\\ K_{5}=
\left(\begin{matrix}
\tilde{\lambda}&\texttt{0}&\texttt{0}&\texttt{0}
\\
\texttt{0}&-\tilde{\lambda}&\texttt{0}&\texttt{0}\\\texttt{0}&\texttt{0}&-\tilde{\lambda}
&\texttt{0}\\\texttt{0}&\texttt{0}
&\texttt{0}&-\tilde{\lambda}\end{matrix}\right).\quad\quad\quad\quad\quad\tag{\ref{e24}}
\end{align}
\end{subequations}

\item{\textsl{Solutions with
$m(X)=X(X-\tilde{\lambda}$$\mathbf{1}_{4})$}}

The roots of the minimum polynomial and the eigenvalues of the
solutions in this case are $\{0, \tilde{\lambda}\}$. They are not
repeated so, the corresponding matrices are diagonalizable. The
Jordan canonical forms with various eigenvalue multiplicities are
\begin{subequations}\label{e24.1}
\begin{align}
 K_6=\left(\begin{matrix}
\tilde{\lambda}&\texttt{0}&\texttt{0}&\texttt{0}
\\
\texttt{0}&\tilde{\lambda}&\texttt{0}&\texttt{0}\\\texttt{0}&\texttt{0}&\tilde{\lambda}
&\texttt{0}\\\texttt{0}&\texttt{0}
&\texttt{0}&\texttt{0}\end{matrix}\right),
K_{7}=\left(\begin{matrix}
\tilde{\lambda}&\texttt{0}&\texttt{0}&\texttt{0}
\\
\texttt{0}&\tilde{\lambda}&\texttt{0}&\texttt{0}\\\texttt{0}&\texttt{0}&\texttt{0}
&\texttt{0}\\\texttt{0}&\texttt{0}
&\texttt{0}&\texttt{0}\end{matrix}\right),\notag\\ K_{8}=
\left(\begin{matrix}
\tilde{\lambda}&\texttt{0}&\texttt{0}&\texttt{0}
\\
\texttt{0}&\texttt{0}&\texttt{0}&\texttt{0}\\\texttt{0}&\texttt{0}&\texttt{0}
&\texttt{0}\\\texttt{0}&\texttt{0}
&\texttt{0}&\texttt{0}\end{matrix}\right).\quad\quad\quad\quad\quad\tag{\ref{e24.1}}
\end{align}
\end{subequations}

\item{\textsl{Solutions with
$m(X)=X(X+\tilde{\lambda}$$\mathbf{1}_{4})$}}

Now, the roots of the minimum polynomial and the eigenvalues of
the solutions become $\{0, -\tilde{\lambda}\}$. Again, they are
distinct, the corresponding matrices are diagonalizable. The
possible Jordan canonical forms with various eigenvalue
multiplicities are
\begin{subequations}\label{e25}
\begin{align}
 K_9=\left(\begin{matrix}
-\tilde{\lambda}&\texttt{0}&\texttt{0}&\texttt{0}
\\
\texttt{0}&-\tilde{\lambda}&\texttt{0}&\texttt{0}\\\texttt{0}&\texttt{0}&-\tilde{\lambda}
&\texttt{0}\\\texttt{0}&\texttt{0}
&\texttt{0}&\texttt{0}\end{matrix}\right),
K_{10}=\left(\begin{matrix}
-\tilde{\lambda}&\texttt{0}&\texttt{0}&\texttt{0}
\\
\texttt{0}&-\tilde{\lambda}&\texttt{0}&\texttt{0}\\\texttt{0}&\texttt{0}&\texttt{0}
&\texttt{0}\\\texttt{0}&\texttt{0}
&\texttt{0}&\texttt{0}\end{matrix}\right),\notag\\ K_{11}=
\left(\begin{matrix}
-\tilde{\lambda}&\texttt{0}&\texttt{0}&\texttt{0}
\\
\texttt{0}&\texttt{0}&\texttt{0}&\texttt{0}\\\texttt{0}&\texttt{0}&\texttt{0}
&\texttt{0}\\\texttt{0}&\texttt{0}
&\texttt{0}&\texttt{0}\end{matrix}\right).\quad\quad\quad\quad\quad\quad
\:\tag{\ref{e25}}
\end{align}
\end{subequations}

\item{\textsl{Solutions with $m(X)=X(AX^2+C\mathbf{1}_{4})$}}

In this case, the roots of the minimum polynomial are $\{0,
\tilde{\lambda}, -\tilde{\lambda}\}$. They are not repeated,
hence, the corresponding matrices are diagonalizable. The possible
Jordan canonical forms constructed with various multiplicities of
these eigenvalues can be listed as
\begin{subequations}\label{e26}
\begin{align}
 K_{12}=\left(\begin{matrix}
\tilde{\lambda}&\texttt{0}&\texttt{0}&\texttt{0}
\\
\texttt{0}&-\tilde{\lambda}&\texttt{0}&\texttt{0}\\\texttt{0}&\texttt{0}&\texttt{0}
&\texttt{0}\\\texttt{0}&\texttt{0}
&\texttt{0}&\texttt{0}\end{matrix}\right),
K_{13}=\left(\begin{matrix}
\tilde{\lambda}&\texttt{0}&\texttt{0}&\texttt{0}
\\
\texttt{0}&\tilde{\lambda}&\texttt{0}&\texttt{0}\\\texttt{0}&\texttt{0}&-\tilde{\lambda}
&\texttt{0}\\\texttt{0}&\texttt{0}
&\texttt{0}&\texttt{0}\end{matrix}\right),\notag\\ K_{14}=
\left(\begin{matrix}
\tilde{\lambda}&\texttt{0}&\texttt{0}&\texttt{0}
\\
\texttt{0}&-\tilde{\lambda}&\texttt{0}&\texttt{0}\\\texttt{0}&\texttt{0}&-\tilde{\lambda}
&\texttt{0}\\\texttt{0}&\texttt{0}
&\texttt{0}&\texttt{0}\end{matrix}\right).\quad\quad\quad\quad\quad\quad\tag{\ref{e26}}
\end{align}
\end{subequations}

\begin{itemize}
\item\title{\textbf{\underline{The cases when $C=0$\:\: :}}}
\end{itemize}

\item{\textsl{Solutions with $m(X)=AX+B\mathbf{1}_{4}$}}

Since, the minimum polynomial is linear this case is a diagonal
solution. It is

\begin{equation}\label{e27}
X_1=-\frac{B}{A}\mathbf{1}_{4}.
\end{equation}

\item{\textsl{Solutions with $m(X)=X(AX+B\mathbf{1}_{4})$}}

The roots of $m(X)$ now, become $\{0,-B/A\}$. They are distinct
thus, the matrices with this minimum polynomial are diagonalizable
with eigenvalues $\{0,-B/A\}$. The possible Jordan normal forms
are

\begin{subequations}\label{e28}
\begin{align}
 X_{2}=\left(\begin{matrix}
-B/A&\texttt{0}&\texttt{0}&\texttt{0}
\\
\texttt{0}&\texttt{0}&\texttt{0}&\texttt{0}\\\texttt{0}&\texttt{0}&\texttt{0}
&\texttt{0}\\\texttt{0}&\texttt{0}
&\texttt{0}&\texttt{0}\end{matrix}\right),
X_{3}=\left(\begin{matrix} -B/A&\texttt{0}&\texttt{0}&\texttt{0}
\\
\texttt{0}&-B/A&\texttt{0}&\texttt{0}\\\texttt{0}&\texttt{0}&\texttt{0}
&\texttt{0}\\\texttt{0}&\texttt{0}
&\texttt{0}&\texttt{0}\end{matrix}\right),\notag\\ X_{4}=
\left(\begin{matrix} -B/A&\texttt{0}&\texttt{0}&\texttt{0}
\\
\texttt{0}&-B/A&\texttt{0}&\texttt{0}\\\texttt{0}&\texttt{0}&-B/A
&\texttt{0}\\\texttt{0}&\texttt{0}
&\texttt{0}&\texttt{0}\end{matrix}\right).\quad\quad\quad\quad\quad\quad\tag{\ref{e28}}
\end{align}
\end{subequations}

\item{\textsl{Solutions with $m(X)=X^2(AX+B\mathbf{1}_{4})$}}

For this branch of solutions, the minimum polynomial $m(X)$ has
repeated roots which are $\{0,0,-B/A\}$. Hence, the matrices with
this minimum polynomial are nondiagonalizable. In the relevant
Jordan canonical forms, the dimension of the secondary block
corresponding to the zero eigenvalue is two, and for the $-B/A$
eigenvalue it is one. The possible Jordan forms are

\begin{subequations}\label{e29}
\begin{align}
 X_{5}=\left(\begin{matrix}
-B/A&\texttt{0}&\texttt{0}&\texttt{0}
\\
\texttt{0}&\texttt{0}&\texttt{1}&\texttt{0}\\\texttt{0}&\texttt{0}&\texttt{0}
&\texttt{0}\\\texttt{0}&\texttt{0}
&\texttt{0}&\texttt{0}\end{matrix}\right),
X_{6}=\left(\begin{matrix} -B/A&\texttt{0}&\texttt{0}&\texttt{0}
\\
\texttt{0}&\texttt{0}&\texttt{0}&\texttt{0}\\\texttt{0}&\texttt{0}&\texttt{0}
&\texttt{1}\\\texttt{0}&\texttt{0}
&\texttt{0}&\texttt{0}\end{matrix}\right),\notag\\ X_{7}=
\left(\begin{matrix} -B/A&\texttt{0}&\texttt{0}&\texttt{0}
\\
\texttt{0}&-B/A&\texttt{0}&\texttt{0}\\\texttt{0}&\texttt{0}&\texttt{0}
&\texttt{1}\\\texttt{0}&\texttt{0}
&\texttt{0}&\texttt{0}\end{matrix}\right).\quad\quad\quad\quad\quad\quad\tag{\ref{e29}}
\end{align}
\end{subequations}
\end{enumerate}
\subsection{The parametric structure}

In the previous subsection, we have derived and classified all the
representatives of the similarity equivalence classes of the
solutions of Eq.\eqref{e12}. The reader may verify that the
matrices listed for various conditions satisfy Eq.\eqref{e12} by
direct substitution for unspecified constant coefficients under
these conditions. Therefore, all the matrix functions constructed
from these representatives by similarity transformations via an
invertible matrix function $P(x^\mu)$ also satisfy Eq.\eqref{e12}.
However, in Eq.\eqref{e10} the coefficients of the matrix equation
for $\sqrt{\Sigma}$ are not numerical constants, on the contrary,
they are functions of the elementary symmetric polynomials
$e_1,e_2$ of $\sqrt{\Sigma}$ which turn this matrix equation into
a very nontrivial one. In this respect, our task is not over and
we have to furthermore solve the entries of the matrices listed in
the previous subsection explicitly, so that they satisfy this
nontrivial matrix equation Eq.\eqref{e10}.

\begin{itemize}
\item\title{\textbf{\underline{The solutions;
$U_{1,2,3,4,5},V_{0,1,2,3,4},Y$\:\: :}}}
\end{itemize}
In this case, the solutions satisfy
\begin{equation}\label{e30}
AX^2+BX+C\mathbf{1}_{4}=0.
\end{equation}
By taking the trace of this equation, and using Eqs.\eqref{e3},
and \eqref{e11} we get
\begin{equation}\label{e31}
3\beta_2e_1+2\beta_3e_2+4\beta_1=0.
\end{equation}
For the cases which satisfy $B^2-4AC>0$ the roots of
Eq.\eqref{e30} are
\begin{equation}\label{e32}
\lambda_1=\frac{-B+\sqrt{B^2-4AC}}{2A},\quad
\lambda_2=\frac{-B-\sqrt{B^2-4AC}}{2A}.
\end{equation}
From Eqs.\eqref{e14}, and \eqref{e15} we have
\begin{equation}\label{e33}
e_1=tr(U_i)=n\lambda_1+m\lambda_2,
\end{equation}
for $U_1,U_2,U_3,U_4,U_5$ the $(n,m)$ values are
$(4,0),(0,4),(2,2),(3,1),(1,3)$, respectively. When $m-n\neq \pm
2$ namely, for $U_1,U_2,U_3$ using Eqs.\eqref{e31}, and
\eqref{e32} in Eq.\eqref{e33} yields the quadratic equation
\begin{equation}\label{e34}
a(e_1)^2+be_1+c=0,
\end{equation}
with
\begin{subequations}\label{e35}
\begin{align}
a&=2(n-1)(m-1)(\beta_3)^2,\notag\\
b&=-\big((m-n)^2+2(m+n-2mn)\big)\beta_2\beta_3,\notag\\
c&=2mn(\beta_2)^2-2(n-m)^2\beta_1\beta_3.\tag{\ref{e35}}
\end{align}
\end{subequations}
Therefore, provided
\begin{equation}\label{e36}
b^2-4ac\geq 0,
\end{equation}
$e_1$ becomes
\begin{equation}\label{e37}
e_1=\frac{-b\pm\sqrt{b^2-4ac}}{2a},
\end{equation}
and from Eq.\eqref{e31} we have
\begin{equation}\label{e38}
e_2=-\frac{3\beta_2}{2\beta_3}\big(\frac{-b\pm\sqrt{b^2-4ac}}{2a}\big)-\frac{2\beta_1}{\beta_3},
\end{equation}
For $U_3$, $b^2-4ac=0$ thus, the condition \eqref{e36} is
automatically satisfied. For $U_1$, and $U_2$ it becomes
\begin{equation}\label{e38.5}
3(\beta_2)^2-4\beta_1\beta_3\geq 0.
\end{equation}
Furthermore, we
can now explicitly write the components of $U_1,U_2,U_3$ as
\begin{equation}\label{e39}
\lambda_1=\frac{\beta_2+\beta_3\big(\frac{-b\pm\sqrt{b^2-4ac}}{2a}\big)-\sqrt{\Delta}}{2\beta_3},\quad
\lambda_2=\frac{\beta_2+\beta_3\big(\frac{-b\pm\sqrt{b^2-4ac}}{2a}\big)+\sqrt{\Delta}}{2\beta_3},
\end{equation}
where
\begin{equation}\label{e40}
\Delta=(\beta_3)^2\big(\frac{-b\pm\sqrt{b^2-4ac}}{2a}\big)^2+4\beta_2\beta_3\big(\frac{-b\pm\sqrt{b^2-4ac}}{2a}\big)+(\beta_2)^2+4\beta_1\beta_3.
\end{equation}
In this expression, by substituting $a,b,c,$ from Eq\eqref{e35}
the restriction $\Delta=B^2-4AC>0$ that defines these solutions
takes the form
\begin{equation}\label{e41}
-3(\beta_2)^2+4\beta_1\beta_3> 0,
\end{equation}
for $U_3$, and it reads
\begin{equation}\label{e41.5}
3(\beta_2)^2-4\beta_1\beta_3> 0,
\end{equation}
for $U_1$, and $U_2$. The condition \eqref{e41.5} is stronger than
\eqref{e38.5} hence, it must be taken as the defining constraint
on the parameter space for the solutions $U_1,U_2$ to exist.
Therefore, when the coefficients $\{\beta_i\}$ satisfy
Eq.\eqref{e41} the similarity equivalence class of $U_3$ (upon the
substitution of the matrix elements via \eqref{e39}) are solutions
of Eq.\eqref{e10}. Also, when the coefficients $\{\beta_i\}$
satisfy Eq.\eqref{e41.5} the similarity equivalence classes of
$U_1$, and $U_2$ again, by using the relevant matrix entries from
\eqref{e39} are solutions of Eq.\eqref{e10}. On the other hand,
when $m-n=\pm 2$, namely, for the cases $U_4,U_5$ Eqs.\eqref{e31},
\eqref{e32}, and \eqref{e33} lead us to the condition
\begin{equation}\label{e42}
3(\beta_2)^2-4\beta_1\beta_3=0.
\end{equation}
By using this condition one can show that we always have
\begin{equation}\label{e43}
\Delta_{m-n=\pm 2}=B^2-4AC=(2\beta_2+\beta_3 e_1)^2>0.
\end{equation}
Thus, for these cases $\Delta>0$ is always satisfied provided
\eqref{e42} holds. The roots of the minimum polynomial now become
\begin{equation}\label{e44}
\lambda_1=\frac{\beta_2+\beta_3e_1-(\pm(2\beta_2+\beta_3e_1))}{2\beta_3},\quad
\lambda_2=\frac{\beta_2+\beta_3e_1+(\pm(2\beta_2+\beta_3e_1))}{2\beta_3}.
\end{equation}
By using these in $U_4$, and $U_5$ the consistency condition
\eqref{e33} demands that the signs in \eqref{e44} must be chosen
such that
\begin{subequations}\label{e45}
\begin{align}
 U_4&=\left(\begin{matrix}
-\frac{\beta_2}{2\beta_3}&\texttt{0}&\texttt{0}&\texttt{0}
\\
\texttt{0}&-\frac{\beta_2}{2\beta_3}&\texttt{0}&\texttt{0}\\\texttt{0}&\texttt{0}&-\frac{\beta_2}{2\beta_3}
&\texttt{0}\\\texttt{0}&\texttt{0}
&\texttt{0}&\frac{3\beta_2+2\beta_3e_1}{2\beta_3}\end{matrix}\right),\notag\\
U_{5}&=\left(\begin{matrix}
\frac{3\beta_2+2\beta_3e_1}{2\beta_3}&\texttt{0}&\texttt{0}&\texttt{0}
\\
\texttt{0}&-\frac{\beta_2}{2\beta_3}&\texttt{0}&\texttt{0}\\\texttt{0}&\texttt{0}&-\frac{\beta_2}{2\beta_3}
&\texttt{0}\\\texttt{0}&\texttt{0}
&\texttt{0}&-\frac{\beta_2}{2\beta_3}\end{matrix}\right),\tag{\ref{e45}}
\end{align}
\end{subequations}
which are similar to each other thus, we will take the
representative of the equivalence class as $U\equiv U_4$. We
remark that, in this case the formulation does not specify $e_1$
and it remains as a free parameter. It can take any value except
$-2\beta_2/\beta_3$ which would violate \eqref{e43}. Therefore,
the solution $U\equiv U_4$, is parametrized by a free trace
parameter $e_1$ as in Eq.\eqref{e45}, and from \eqref{e31} we have
\begin{equation}\label{e46}
e_2=-\frac{3\beta_2}{2\beta_3}e_1-\frac{2\beta_1}{\beta_3}.
\end{equation}
Among the solutions which satisfy Eq.\eqref{e30}; when $B^2-4AC=0$
we have the ones $V_0,V_1,V_2,V_3,V_4$, and when $B^2-4AC<0$ we
have the solution $Y$. For all of these cases taking the trace of
these matrices leads us to
\begin{equation}\label{e47}
e_1=-\frac{2\beta_2}{\beta_3}.
\end{equation}
From Eq.\eqref{e31} we get
\begin{equation}\label{e48}
e_2=\frac{3(\beta_2)^2}{(\beta_3)^2}-\frac{2\beta_1}{\beta_3}.
\end{equation}
From these relations for $V_0,V_1,V_2,V_3,V_4$ via the definitions
in Eq\eqref{e11} we have
\begin{equation}\label{e49}
\lambda^\prime=-\frac{B}{2A}=-\frac{\beta_2}{2\beta_3},
\end{equation}
and the condition $B^2-4AC=0$ which allows these solutions to
exist becomes
\begin{equation}\label{e50}
3(\beta_2)^2-4\beta_1\beta_3=0.
\end{equation}
On the other hand, when $B^2-4AC<0$ is valid Eq.\eqref{e30} has
two complex roots $\lambda=R+Ii$, and $\lambda^*=R-Ii$, here
\begin{equation}\label{e51}
R=-\frac{\beta_2}{2\beta_3},\quad
I=-\frac{\sqrt{3(\beta_2)^2-4\beta_1\beta_3}}{2\beta_3},
\end{equation}
where we have used Eqs.\eqref{e11}, \eqref{e47}, and \eqref{e48}.
From the same equations the condition $B^2-4AC<0$ which allows the
solution $Y$ to exist now, becomes
\begin{equation}\label{e52}
-3(\beta_2)^2+4\beta_1\beta_3<0.
\end{equation}
\begin{itemize}
\item\title{\textbf{\underline{The solutions;
$U_{6,...,14},V_{5,...,10},Y^\prime$\:\: :}}}
\end{itemize}
For the $U-$matrices
\begin{equation}\label{e53}
e_1=n\lambda_1+m\lambda_2=-(n+m)\frac{B}{2A}+(n-m)\frac{\sqrt{B^2-4AC}}{2A},
\end{equation}
where
$(n,m)=(1,0),(2,0),(3,0),(0,1),(0,2),(0,3),(2,1),(1,1),(1,2)$ for
$U_6,U_7,U_8,U_9,U_{10},U_{11},U_{12},U_{13},U_{14},$
respectively. We also have
\begin{equation}\label{e54}
e_2=\frac{1}{2}((e_1)^2-tr(U_i)^2).
\end{equation}
Now, from this relation by using
$tr(U_i)^2=n(\lambda_1)^2+m(\lambda_2)^2$, and the fact that
$\lambda_1,\lambda_2$ are the roots of $Ax^2+Bx+C=0$ we obtain
\begin{equation}\label{e55}
e_2=\frac{1}{(2-n-m)}\big((n+m-1)\frac{\beta_2}{\beta_3}e_1+(n+m)\frac{\beta_1}{\beta_3}\big),
\end{equation}
when $n+m\neq 2$. Again, for the cases $n+m\neq 2$ substituting
Eqs. \eqref{e11}, and \eqref{e55} into Eq.\eqref{e53} after some
algebra leads us to the equation
\begin{equation}\label{e56}
a^\prime(e_1)^2+b^\prime e_1+c^\prime=0,
\end{equation}
where
\begin{subequations}\label{e57}
\begin{align}
a^\prime&=(2-n-m)(1-m)(1-n)(\beta_3)^2,\notag\\
b^\prime&=-\big((2-n-m)(1-m)n+(2-n-m)(1-n)m-(n-m)^2\big)\beta_2\beta_3,\notag\\
c^\prime&=mn(2-n-m)(\beta_2)^2+2(n-m)^2\beta_1\beta_3.\tag{\ref{e57}}
\end{align}
\end{subequations}
If the condition
\begin{equation}\label{e58}
b^{\prime 2}-4a^\prime c^\prime\geq 0,
\end{equation}
is satisfied then $e_1$ becomes
\begin{equation}\label{e59}
e_1=\frac{-b^\prime\pm\sqrt{b^{\prime 2}-4a^\prime
c^\prime}}{2a^\prime},
\end{equation}
and via  Eq.\eqref{e55} we get
\begin{equation}\label{e60}
e_2=\frac{1}{(2-n-m)}\big((n+m-1)\frac{\beta_2}{\beta_3}\big(\frac{-b^\prime\pm\sqrt{b^{\prime
2}-4a^\prime c^\prime}}{2a^\prime}\big
)+(n+m)\frac{\beta_1}{\beta_3}\big),
\end{equation}
which is now written solely in terms of the parameters
$\{\beta_i\}$'s. We can also write the components of the
$U$-matrices as
\begin{equation}\label{e61}
\lambda_{1,2}=-\frac{1}{2\beta_3}\bigg(-\big(\beta_2+\beta_3\big(\frac{-b^\prime\pm\sqrt{b^{\prime
2}-4a^\prime
c^\prime}}{2a^\prime}\big)\big)\pm\sqrt{\Delta}\bigg),
\end{equation}
where
\begin{subequations}\label{e62}
\begin{align}
\Delta&=B^2-4AC\notag\\
&=\bigg(\beta_2+\beta_3\big(\frac{-b^\prime\pm\sqrt{b^{\prime
2}-4a^\prime
c^\prime}}{2a^\prime}\big)\bigg)^2+4\beta_3\bigg[-\beta_1-\beta_2\big(\frac{-b^\prime\pm\sqrt{b^{\prime
2}-4a^\prime
c^\prime}}{2a^\prime}\big)\notag\\
&\quad-\frac{1}{(2-n-m)}\bigg((n+m-1)\beta_2\big(\frac{-b^\prime\pm\sqrt{b^{\prime
2}-4a^\prime c^\prime}}{2a^\prime}\big
)+(n+m)\beta_1\bigg)\bigg]\tag{\ref{e62}}.
\end{align}
\end{subequations}
For the existence of these solutions we must have $\Delta>0$,
together with the condition \eqref{e58}. In spite of the fact
that, Eq.\eqref{e56} is generally valid for all of the matrices;
$U_{6,8,9,11,12,14}$ we must be cautious since for some of these
solutions it reduces to a redundant form, or it is trivially
satisfied. We will consider such cases one by one. In particular,
for $U_6$, and $U_9$ \eqref{e58} is satisfied directly as for
these cases $b^{\prime 2}-4a^\prime c^\prime=0$. However, for
these cases Eq.\eqref{e56} holds only if $\beta_1\beta_3=0$, thus,
we will exclude the solutions $U_6,U_9$. For $U_8,U_{11}$
\eqref{e58} becomes $(\beta_2)^2-\beta_1\beta_3\geq 0$, and
$\Delta>0$ leads to
\begin{equation}\label{e63}
(\beta_2)^2-\beta_1\beta_3>0,
\end{equation}
which is stronger than \eqref{e58}. Therefore, for the existence
of the solutions $U_8,U_{11}$, Eq.\eqref{e63} is the only
condition. For the matrices $U_{12},U_{14}$  we have $b^{\prime
2}-4a^\prime c^\prime=0$ hence, Eq.\eqref{e58} is satisfied. Also,
now Eq.\eqref{e56} turns into the condition
\begin{equation}\label{e64}
-(\beta_2)^2+\beta_1\beta_3=0,
\end{equation}
and by using this condition for these cases, we find that
\begin{equation}\label{e64.5}
\Delta=(3\beta_2+\beta_3 e_1)^2>0.
\end{equation}
We observe that, for $U_{12},U_{14}$ if \eqref{e64} is satisfied
we always have $\Delta>0$, this fact makes Eq.\eqref{e64} the
unique condition for the existence of these solutions. For the
cases $U_{12},U_{14}$ the roots of the minimum polynomial are
\begin{equation}\label{e65}
\lambda_1=\frac{\beta_2+\beta_3e_1-(\pm(3\beta_2+\beta_3e_1))}{2\beta_3},\quad
\lambda_2=\frac{\beta_2+\beta_3e_1+(\pm(3\beta_2+\beta_3e_1))}{2\beta_3}.
\end{equation}
When we substitute them in the matrices $U_{12}$, and $U_{14}$ the
consistency condition \eqref{e53} constrains us to choose the
signs in Eq.\eqref{e65} in such a way that
\begin{equation}\label{e66}
 U_{12}=U_{14}\equiv U^{\prime}=\left(\begin{matrix}
\texttt{0}&\texttt{0}&\texttt{0}&\texttt{0}
\\
\texttt{0}&-\frac{\beta_2}{\beta_3}&\texttt{0}&\texttt{0}\\\texttt{0}&\texttt{0}&-\frac{\beta_2}{\beta_3}
&\texttt{0}\\\texttt{0}&\texttt{0}
&\texttt{0}&\frac{2\beta_2+\beta_3e_1}{\beta_3}\end{matrix}\right),
\end{equation}
where $e_1$ remains to be a free trace parameter, as the matrix
\eqref{e66} trivially satisfies Eq.\eqref{e53}. It can take any
value except $-3\beta_2/\beta_3$ which would violate
\eqref{e64.5}. For the solution $U^\prime$, which is parametrized
by $e_1$ from Eq.\eqref{e55} we have
\begin{equation}\label{e67}
e_2=-\frac{2\beta_2}{\beta_3}e_1-\frac{3\beta_1}{\beta_3}.
\end{equation}
On the other hand, if we turn our attention to the cases when
$n+m=2$, namely, the cases; $U_7,U_{10},U_{13}$ again, via
$tr(U_i)^2=n(\lambda_1)^2+m(\lambda_2)^2$, and the fact that
$\lambda_1,\lambda_2$ satisfy $Ax^2+Bx+C=0$, Eq.\eqref{e54} gives
\begin{equation}\label{e68}
e_1=-\frac{2\beta_1}{\beta_2}.
\end{equation}
For $U_{13}$, we have $e_1=\lambda_1+\lambda_2$ however, this
relation together with Eq.\eqref{e68} result in  the constraint
$\beta_2=0$. From Eq.\eqref{e68} in this case we must also have
$\beta_1=0$. Therefore, we will disregard $U_{13}$. For the
matrices $U_7,U_{10}$,
\begin{equation}\label{e69}
e_1=2\lambda_{1,2}=2\bigg(\frac{-B\pm\sqrt{B^2-4AC}}{2A}\bigg),
\end{equation}
where $+$ corresponds to $U_7$, and $-$ to $U_{10}$, respectively.
By referring to the definitions in Eq.\eqref{e11}, and by using
Eq.\eqref{e68} in Eq.\eqref{e69} we find that
\begin{equation}\label{e70}
e_2=\frac{(\beta_1)^2}{(\beta_2)^2}.
\end{equation}
For the solutions $U_7,U_{10}$, upon the substitution of $e_1,e_2$
from Eqs.\eqref{e68}, and \eqref{e70} the defining condition
$B^2-4AC>0$ becomes
\begin{equation}\label{e71}
(\beta_2)^2>0,
\end{equation}
which is automatically satisfied. For these cases, by referring to
the Eqs. \eqref{e11}, \eqref{e68}, and \eqref{e70} the computation
of the matrix entries reads
\begin{equation}\label{e72}
\lambda_{1,2}=\frac{(\beta_2)^2-2\beta_1\beta_3\mp(\pm(\beta_2)^2)}{2\beta_3\beta_2}.
\end{equation}
By substituting these in the matrices $U_{7}$, and $U_{10}$, the
consistency condition \eqref{e68} denotes that the signs in
\eqref{e72} must be chosen such that
\begin{equation}\label{e73}
 U_{7}=U_{10}\equiv U^{\prime\prime}=\left(\begin{matrix}
-\frac{\beta_1}{\beta_2}&\texttt{0}&\texttt{0}&\texttt{0}
\\
\texttt{0}&-\frac{\beta_1}{\beta_2}&\texttt{0}&\texttt{0}\\\texttt{0}&\texttt{0}&\texttt{0}
&\texttt{0}\\\texttt{0}&\texttt{0}
&\texttt{0}&\texttt{0}\end{matrix}\right).
\end{equation}
Let us now focus on the solutions $V_{5,6,7,8,9,10}$. They have
\begin{equation}\label{e74}
e_1=trV_i=n\lambda^\prime,\quad
tr(V_i)^2=n(\lambda^\prime)^2=\frac{(e_1)^2}{n},
\end{equation}
where for $V_5,V_6,V_{10}$ $n=3$, for $V_7,V_9$ $n=2$, and for
$V_8$ $n=1$. Furthermore,
\begin{equation}\label{e75}
e_2=\frac{1}{2}\big((e_1)^2-tr(V_i)^2\big)=\frac{(n-1)}{2n}(e_1)^2.
\end{equation}
Referring to the definitions of the coefficients in Eq.\eqref{e11}
once more, explicitly we have
\begin{equation}\label{e76}
\lambda^\prime=-\frac{B}{2A}=\frac{\beta_2+\beta_3 e_1}{2\beta_3}.
\end{equation}
Using this relation in Eq.\eqref{e74} gives
\begin{equation}\label{e77}
e_1=\frac{n\beta_2}{(2-n)\beta_3},
\end{equation}
when $n\neq 2$. If we consider the matrices $V_7,V_9$ for which
$n=2$ this computation leads to $\beta_2=0$, hence, we also
disregard these solutions. Next, substituting Eq.\eqref{e77} in
Eq.\eqref{e75} we also obtain $e_2$ in terms of the
$\beta_i$-parameters as
\begin{equation}\label{e78}
e_2=\frac{n(n-1)(\beta_2)^2}{2(2-n)^2(\beta_3)^2}.
\end{equation}
Finally, by substituting Eq.\eqref{e77} into Eq.\eqref{e76} we can
also calculate explicitly the matrix entries as
\begin{equation}\label{e79}
\lambda^\prime=\frac{\beta_2}{(2-n)\beta_3}.
\end{equation}
The condition $B^2-4AC=0$, needed for the existence of the
solutions $V_{5,6,7,8,9,10}$ reads
\begin{equation}\label{e80}
\bigg[\frac{1}{(2-n)^2}-\frac{n}{2-n}-\frac{n(n-1)}{2(n-2)^2}\bigg](\beta_2)^2-\beta_1\beta_3=0,
\end{equation}
where we have used Eqs. \eqref{e11}, \eqref{e77}, and \eqref{e78}.
In particular, for the case $V_8$; $n=1$, this condition gives
$\beta_1\beta_3=0$ thus, we also disregard $V_8$. For the
remaining cases $V_{5,6,10}$, since $n=3$ Eq.\eqref{e80} takes the
form
\begin{equation}\label{e81}
(\beta_2)^2-\beta_1\beta_3=0.
\end{equation}
Our final case in this set of solutions is $Y^\prime$ which exists
when $B^2-4AC<0$. Starting from its definition in Eq.\eqref{e23}
we find that
\begin{equation}\label{e82}
e_1=trY^\prime=2R,
\end{equation}
where
\begin{equation}\label{e83}
R=-\frac{B}{2A}.
\end{equation}
By substituting the definitions of $A,B$ from  Eq.\eqref{e11} into
Eq.\eqref{e82} we get $\beta_2=0$. Therefore, $Y^\prime$ will also
be excluded from our solution space which will be constructed for
nonzero $\beta_i-$coefficients in Eq.\eqref{e1}.
\begin{itemize}
\item\title{\textbf{\underline{The solutions;
$X_{1,...,7},K_{1,...,14}$\:\: :}}}
\end{itemize}
In general, we will exclude the cases when $A=0$ as they
correspond to $\beta_3=0$. The cases when $B=0$ with $C/A>0$ may
have the minimum polynomials either $m(X)=AX^2+C\mathbf{1}_4$, or
$m(X)=X(AX^2+C\mathbf{1}_4)$. However, both cases are restricted
to $\beta_2=0$. Thus, we will exclude them. On the other hand,
when $B=0$, $C/A<0 $ we have the solutions $K_{1,...,14}$. Here
\begin{equation}\label{e84}
e_1=tr(K_j)=n\tilde{\lambda}=n\sqrt{-C/A},
\end{equation}
and
\begin{equation}\label{e85}
tr(K_j)^2=m(\tilde{\lambda})^2=\frac{m}{n^2}(e_1)^2,
\end{equation}
where $(n,m)=$$(4,4)$, $(-4,4)$, $(2,4)$, $(0,4)$, $(-2,4)$,
$(3,3)$, $(2,2)$, $(1,1)$, $(-3,3)$, $(-2,2)$, $(-1,1)$, $(0,2)$,
$(1,3)$, $(-1,3)$ for $K_{1,2,...,13,14}$, respectively. For
$K_4,K_{12}$, $e_1=0$, from the condition $B=0$ we observe that
for these cases to be solutions we must have $\beta_2=0$ thus, we
exclude these solutions too. For the other cases from the above
relations we have
\begin{equation}\label{e86}
e_2=\frac{1}{2}\big((e_1)^2-tr(K_j)^2\big)=\frac{n^2-m}{2n^2}(e_1)^2.
\end{equation}
Now, if we use Eqs. \eqref{e11}, and \eqref{e86} in Eq.\eqref{e84}
we obtain the quadratic equation
\begin{equation}\label{e87}
\big(1+\frac{n^2-m}{2}\big)(e_1)^2+\frac{n^2\beta_2}{\beta_3}e_1+\frac{n^2\beta_1}{\beta_3}=0.
\end{equation}
When $n^2-m\neq -2$ this equation has the solutions
\begin{equation}\label{e88}
e_1=\frac{-\frac{n^2\beta_2}{\beta_3}\pm
\sqrt{\frac{n^4(\beta_2)^2}{(\beta_3)^2}-2\big(2+n^2-m\big)\frac{n^2\beta_1}{\beta_3}}}{2+n^2-m}.
\end{equation}
We observe that if
\begin{equation}\label{e89}
\frac{n^2(\beta_2)^2}{(\beta_3)^2}-2\big(2+n^2-m\big)\frac{\beta_1}{\beta_3}\geq
0,
\end{equation}
$e_1$ is real and via the relation Eq.\eqref{e84} the condition
$C/A<0$ is satisfied. For these solutions to exist we must also
satisfy the condition $B=0$, which reads
\begin{equation}\label{e90}
\big[n^4-(m-2)^2\big](\beta_2)^2-2\big(2+n^2-m\big)n^2\beta_1\beta_3=0,
\end{equation}
where we have made use of the Eqs.\eqref{e11}, and \eqref{e88}.
For the solutions $K_{8},K_{11}$ \eqref{e90} gives
$\beta_1\beta_3=0$, which leads us to exclude these solutions. A
close inspection denotes that when \eqref{e90} is satisfied
\eqref{e89} is automatically satisfied, thus, Eq.\eqref{e90} is a
stronger condition to be considered uniquely. Therefore, the
condition \eqref{e90} is the unique defining condition for the
existence of the solutions
$K_{1},K_{2},K_{3},K_{5},K_{6},K_{7},K_{9},K_{10}$. On the other
hand, when $n^2-m=-2$, namely, for the cases $K_{13},K_{14}$
Eq.\eqref{e87} leads us to the result
\begin{equation}\label{e91}
e_1=-\frac{\beta_1}{\beta_2}.
\end{equation}
$e_2$ must again be read from Eq.\eqref{e86}. In this case, since
$\tilde{\lambda}$ must be positive, via Eq.\eqref{e84} we see
that; when $\beta_1/\beta_2>0$, the solution is $K_{14}$, and when
$\beta_1/\beta_2<0$, the solution is $K_{13}$. Besides, since
$C/A=-n^{-2}(e_1)^2=-(e_1)^2$ the condition $C/A<0$ is
automatically satisfied for these solutions. The other restriction
$B=\beta_2+\beta_3e_1=0$ becomes
\begin{equation}\label{e92}
(\beta_2)^2-\beta_1\beta_3=0,
\end{equation}
which is left as the unique defining condition for $K_{13},K_{14}$
to exist. For all these solutions discussed above, the matrix
entries can be found as $\tilde{\lambda}=\sqrt{-C/A}=e_1/n$ by
substituting the relevant $e_1$, and $n$ values case by case. We
should state however, that when one computes $e_1$ from
Eq.\eqref{e88}, then substitutes the results in Eqs.
\eqref{e23.5}, \eqref{e24.1}, \eqref{e25} for the cases $K_{1},$
$K_{2}$, $K_{6},$ $K_{7}$, $K_{9},$ and $K_{10}$, respectively one
finds that $K_1=K_2$, $K_6=K_9$, and $K_7=K_{10}$. For this
reason, we will define the matrices $K\equiv K_1=K_2$,
$K^{\prime}\equiv K_6=K_9$, and $K^{\prime\prime}\equiv
K_7=K_{10}$ for the rest of our analysis. Now, we will turn our
attention to the condition $C=0$ which generates the $X-$series of
solutions. Firstly we should note that when $C=0$, in the previous
subsection, the solutions coming from the minimum polynomials
$m(X)=X$, and $m(X)=X^2$ are excluded as they lead to the
restriction $\beta_1=0$. Besides, the solutions with $m(X)=AX^3$
when $B=C=0$ are also disregarded as they are restricted to the
cases $\beta_1=\beta_2=0$. For $X_1$ we have $e_1=trX_1=-4B/A$
which gives
\begin{equation}\label{e93}
e_1=-\frac{4\beta_2}{3\beta_3}.
\end{equation}
We also get
\begin{equation}\label{e94}
e_2=\frac{1}{2}\big((e_1)^2-tr(X_1)^2\big)=\frac{3}{8}(e_1)^2.
\end{equation}
By using these results, and by referring to Eq.\eqref{e11} once
more, the existence  condition $C=0$ becomes
\begin{equation}\label{e95}
2(\beta_2)^2-3\beta_1\beta_3=0.
\end{equation}
Then, by making use of Eq.\eqref{e93} we can also explicitly write
$X_1$ as
\begin{equation}\label{e96}
X_1=-\frac{\beta_2}{3\beta_3}\textbf{1}_4.
\end{equation}
For the other cases; $X_{2,3,4,5,6,7}$, $e_1=trX_i=-nB/A$ with
$n=1$ for $X_2,X_{5},X_{6}$, and $n=2$ for $X_3,X_7$, and $n=3$
for $X_4$, respectively. Thus, for these cases when $n\neq 1$ we
obtain
\begin{equation}\label{e97}
e_1=-\frac{n\beta_2}{(n-1)\beta_3}.
\end{equation}
For $n=1$, substitution of $A,B$ via Eq.\eqref{e11} in $e_1=-nB/A$
gives $\beta_2=0$, hence, we exclude the solutions
$X_{2},X_5,X_6$. For the solutions $X_{3,4,7}$, we can compute
$e_2$ as
\begin{equation}\label{e98}
e_2=\frac{1}{2}\big((e_1)^2-tr(X_{i})^2\big)=\frac{n-1}{2n}(e_1)^2.
\end{equation}
Similarly, for these cases the condition $C=0$ reads
\begin{equation}\label{e99}
n(\beta_2)^2-2(n-1)\beta_1\beta_3=0.
\end{equation}
Also, by using Eqs.\eqref{e11}, and \eqref{e97} we can explicitly
compute the matrix entries of $X_{3,4,7}$ as
\begin{equation}\label{e100}
-\frac{B}{A}=-\frac{\beta_2}{(n-1)\beta_3}.
\end{equation}
\section{The decoupling solution space}
In the previous section, we have explicitly worked out and
constructed the Jordan canonical forms of the nontrivial
fixed-eigenvalue solutions of Eq.\eqref{e12} in terms of the
$\beta_i-$coefficients when neither of $\beta_{1,2,3}$ is chosen
to be vanishing. Let us first define the set composed of these
solutions
\begin{subequations}\label{e101}
\begin{align}
\mathcal{J}\equiv
\bigg\{&U_1,U_2,U_3,U_8,U_{11},U,U^\prime,U^{\prime\prime},V_0,V_1,V_2,V_3,V_4,V_5,V_6,V_{10},Y,X_{1},\notag\\
&X_{3},X_{4},X_{7},K,K^{\prime},K^{\prime\prime},
K_{3},K_{5},K_{13},K_{14}\bigg\}\tag{\ref{e101}}.
\end{align}
\end{subequations}
In the Appendix, we give a summary of the exact forms of these
matrices whose entries are derived in terms of the
$\beta_i-$coefficients in Section three. Next, we define the set
of matrices
\begin{equation}\label{e101.1}
\mathcal{M}\equiv\bigg\{M_J=M^{-1}JM\:\: \big|\:\: \forall J \in
\mathcal{J},\: \text{and}\:\: \text{det}M\neq 0\bigg\},
\end{equation}
where $M$ is any invertible real constant matrix. As we discussed
before, the similarity equivalence classes of the elements of
$\mathcal{J}$ are also solutions of Eq.\eqref{e10}. Bearing this
in mind now, the set of matrix functions
\begin{equation}\label{e102}
\mathcal{S}\equiv\bigg\{\sqrt{\Sigma}:
\mathcal{U}\rightarrow\mathcal{M}\:\:\big|\:\:  \sqrt{\Sigma}\in
C^\infty\bigg\},
\end{equation}
where $\mathcal{U}$ is a coordinate chart on the $4D$ spacetime
manifold, is the complete set of general nontrivial solutions of
Eq.\eqref{e10} when $\beta_{1,2,3}\neq 0$. Of course, when the
$\beta_i$-parameters are determined the appropriate elements of
$\mathcal{J}$ must be chosen in other words, the ones whose domain
of validity is satisfied by these parameters. We can also define
the subset of $\mathcal{S}$
\begin{equation}\label{e103}
\mathcal{S}\supset\mathcal{P}\equiv\bigg\{\sqrt{\Sigma}\equiv
P_J=P^{-1}(x^\mu)JP(x^\mu)\:\: \big|\:\: \forall J \in
\mathcal{J},\: \text{and}\:\: \text{det}P\neq 0\bigg\},
\end{equation}
where $P(x^\mu)$ is any smooth, invertible real matrix function on
$\mathcal{U}$. The elements of $\mathcal{P}$ have fixed
eigenvalues over the coordinate chart $\mathcal{U}$, and
$\mathcal{P}$ is composed of the matrix functions whose ranges are
grouped in the similarity equivalence classes of the elements of
$\mathcal{J}$. Once more, when constructing $\mathcal{P}$ one must
choose the elements of $\mathcal{J}$ which exist within the domain
defined by the determined $\beta_i$-parameters, thus, one must
respect the validity domains. We should remind the reader of the
fact that the eigenvalues, $e_1$, and $e_2$ are the same for the
elements of a similarity equivalence class. Referring to
Eq.\eqref{e3.1} we now have
\begin{equation}\label{e104}
f=g\Sigma.
\end{equation}
In constructing the sets $\mathcal{S}$, and $\mathcal{P}$ we
treated $\sqrt{\Sigma}$ as a general solution of the matrix
equation Eq.\eqref{e10} without considering its relevance to $g,$
and $f$. However, Eq.\eqref{e104} now brings a constraint on it.
Since $g$, and $f$ are symmetric $\sqrt{\Sigma}$ is restricted to
the condition
\begin{equation}\label{e104.1}
g\Sigma=\Sigma^Tg.
\end{equation}
In other words, not all the elements of $\mathcal{S}$, and
$\mathcal{P}$ will result in a symmetric $f$ in Eq.\eqref{e104}
when a foreground metric $g$ is specified. One, has to consider
the subsets of them whose elements fall into the range of the
product of two symmetric matrices when squared. For the elements
of $\mathcal{P}$, Eq.\eqref{e104.1} becomes
\begin{equation}\label{e104.2}
gP^{-1}J^2P=(P^{-1}J^2P)^Tg.
\end{equation}
When $g$ is specified, one can choose ten function components of
$P(x^\mu)$ freely, and determine the other six in terms of these,
and the components of $g$ by solving the six algebraic equations
arising from Eq.\eqref{e104.2}. Then, one can explicitly construct
the symmetric solution $f$ via Eq.\eqref{e104} in terms of $g$.
Thus, we observe that we have $P=P[g,x^\mu]$ indeed. We should
remark that, although this is the general solution construction
method, one may prefer to choose simplified, or well-designed
forms of $P$ to generate solutions directly. For example, if one
chooses $P=\mathbf{1}_4$, and take a diagonal $g$ one immediately
obtains diagonal $f$ solutions for the choice of the diagonal
elements of $\mathcal{J}$. We now define
\begin{equation}\label{e104.3}
\Gamma_\mathcal{S}=\big\{ (g,f)\:\:\big|\:\: f=gX^2\:\:\big|
X\in\mathcal{S},\:\text{and}\: gX^2=(X^T)^2g\big\},
\end{equation}
and
\begin{equation}\label{e105}
\Gamma_\mathcal{P}=\big\{ (g,f)\:\:\big|\:\: f=g(P_J)^2\:\:\big|
P_J\in\mathcal{P},\:\text{and}\: g(P_J)^2=((P_J)^T)^2g\big\},
\end{equation}
in which, one must construct $P[g,x^\mu]$ in $P_J=P^{-1}JP$ such
that it satisfies Eq\eqref{e104.2} as we discussed above. Next, we
will compute the effective cosmological constant in
Eq.\eqref{e8.3} which enters into the $g$-equation Eq.\eqref{e4}
as an effective presence of the background metric. Taking the
trace of Eq.\eqref{e10} leads us to
\begin{equation}\label{e106}
\beta_1 e_1+2\beta_2 e_2+3\beta_3 e_3=0,
\end{equation}
where we have made use of Eqs.\eqref{e3}. By using this relation
in Eq.\eqref{e8.2}, and by also referring to Eq.\eqref{e2} we
conclude that
\begin{equation}\label{e107}
\tilde{\Lambda}=-\frac{4}{3}\beta_1 e_1-\frac{2}{3}\beta_2 e_2,
\end{equation}
where for a particular solution of the form
\begin{equation}\label{e108}
f=g\big(P^{-1}[g,x^\mu]JP[g,x^\mu]\big)^2=g
P^{-1}[g,x^\mu]J^2P[g,x^\mu],
\end{equation}
with $J\in\mathcal{J}$ one must read the appropriate
$e_1(\beta_i)$, and $e_2(\beta_i)$ for the particular choice of
$J$ from its parametric structure derived in the previous section.
This is due to the fact that $e_1$, and $e_2$ are the symmetric
polynomials of $\sqrt{\Sigma}=P^{-1}[g,x^\mu]JP[g,x^\mu]$, and
under similarity transformations they remain the same so they are
also the symmetric polynomials of the particular $J$ which is
chosen to generate the solution. Since, $\tilde{\Lambda}$ must
stay constant over a coordinate chart $\mathcal{U}$ to be able to
satisfy Eq.\eqref{e8.1}. We will define
\begin{equation}\label{e108.1}
\Gamma=\Gamma_{\mathcal{S}}^\prime\cup\Gamma_{\mathcal{P}},
\end{equation}
as the general solution space of Eq.\eqref{e8} where the set
$\Gamma_{\mathcal{S}}^\prime$ formally corresponds to the elements
in the set $\Gamma_{\mathcal{S}}-\Gamma_{\mathcal{P}}$ with
invariant $e_1,e_2$, or invariant $\tilde{\Lambda}$ over
$\mathcal{U}$. We conclude that, the elements of $\Gamma$
constitute the decoupling solution space of bigravity, with
$\Gamma_{\mathcal{P}}$ being an analytically expressible subset of
it that is composed of similarity equivalence classes of matrix
functions.
\section{Proportional Backgrounds}
Before we conclude, in this section, we present a class of exact
solutions in which the two metrics are proportional to each other
so that
\begin{equation}\label{e110}
f_{\mu\nu}=\mathcal{C}^2g_{\mu\nu}.
\end{equation}
These solutions are generated by the elements of Eq. \eqref{e101}
which are proportional to the unit matrix, namely, by
$\{U_1,U_2,V_0,K,X_1\}$. Among such solutions which are known as
the proportional backgrounds in the literature \cite{8}, as a
special class there exists the subset in which both of the metrics
are diagonal in the same coordinate system. This is due to the
fact that in the action Eq. \eqref{e1} both metrics are defined on
the same coordinate patch, and in our formalism of deriving these
solutions we have kept the same coordinate chart for both of the
metrics all through our analysis. After describing the general
structure of the solutions in Eq. \eqref{e110} we will explicitly
construct examples of bidiagonal-metric solutions in which both of
the metrics are of Friedmann-Lemaitre-Robertson-Walker (FLRW)
type. The values of the proportionality constants $\mathcal{C}$ in
Eq. \eqref{e110} which only depend on the $\{\beta_i\}$-coupling
constants for $\{U_1,U_2,V_0,X_1\}$ are given in the Eqs.
\eqref{e39}, \eqref{e49}, \eqref{e96}, respectively. Also, we have
$\mathcal{C}=\tilde{\lambda}=e_1/n$, with $n=4,-4$ for $K_1,K_2$.
Here, when we computed $e_1$ via Eq. \eqref{e88} for
$(n,m)=(4,4),(-4,4)$ for $K_1,K_2$, respectively we previously
found that $K_1=K_2$, and we have already defined the matrix
$K\equiv K_1=K_2$. Thus, one can take
$\mathcal{C}=\tilde{\lambda}=e_1[K_1]/4$ for $K$. The
proportionality constant $\mathcal{C}$ corresponding to $K$ also
depends only on the coefficients $\{\beta_i\}$. As we mentioned in
Section two, since the configuration in Eq. \eqref{e110} satisfies
Eq. \eqref{e8} the second of the Bianchi identities in Eq.
\eqref{e7.6} is directly fulfilled, and the first one leads to a
contribution of a cosmological constant term to the $g$-metric
equations Eq. \eqref{e4}. The value of this effective cosmological
constant can be computed from  Eq. \eqref{e107} as
\begin{equation}\label{e111}
\tilde{\Lambda}=-\frac{16}{3}\beta_1\mathcal{C}
-4\beta_2\mathcal{C}^2.
\end{equation}
On the other hand, for the solutions of type Eq. \eqref{e110}
since $\tau=0$ we see via Eq. \eqref{e6} that the $f$-metric
equation Eq. \eqref{e4.1} gets no contribution from the $g$-$f$
interaction terms in the action Eq. \eqref{e1}. Therefore, for the
proportional background solutions the $g$, and $f$-metric
equations become
\begin{equation}\label{e112}
G^g_{\mu\nu}-\frac{1}{2}(\Lambda^g+m^2\tilde{\Lambda})
g_{\mu\nu}=8\pi GT^{g}_{M\,\mu\nu},
\end{equation}
and
\begin{equation}\label{e113}
\kappa\big[G^f_{\mu\nu}-\frac{1}{2}\Lambda^f
f_{\mu\nu}\big]=\epsilon 8\pi GT^{f}_{M\,\mu\nu},
\end{equation}
respectively, where $G^g_{\mu\nu}$, and $G^f_{\mu\nu}$ are the
corresponding Einstein tensors. Since, for the solutions Eq.
\eqref{e110} we have
$(G^f)^{\mu}_{\:\:\:\nu}=(G^g)^{\mu}_{\:\:\:\nu}/\mathcal{C}^2$ if
one chooses the fine-tuning of the $g$, and the $f$-sector sources
as
\begin{equation}\label{e114}
\Lambda^f=\frac{1}{\mathcal{C}^2}(\Lambda^g+m^2\tilde{\Lambda}),\quad
(T_M^f)^{\mu}_{\:\:\:\nu}=\frac{\kappa}{\epsilon\,\mathcal{C}^2}
(T_M^g)^{\mu}_{\:\:\:\nu},
\end{equation}
then the equations \eqref{e112}, \eqref{e113} become equivalent
upon the choice of the solutions in Eq. \eqref{e110}. If one of
these equations is satisfied the other is also automatically
satisfied. The fine-tuning that is introduced above is a typical
characteristic of the proportional backgrounds \cite{8}. However,
we should state that the proportional backgrounds derived here are
new and different from the already-known ones in the literature.
The reader may refer to \cite{8} to observe that the known
solutions of type \eqref{e110} are derived by excluding the
possibility of the vanishing of $\tau$ which results in a
different fine-tuning condition that determines the permissible
values of $\mathcal{C}$. Therefore, we conclude that for any
solution of the Einstein equations Eq. \eqref{e112} which are
modified by an effective cosmological constant that is
proportional to $m^2$ with the metric $g$, matter cosmological
constant $\Lambda^g$, and the matter energy-momentum tensor
$T^{g}_{M\,\mu\nu}$ the metric $f$ chosen as in Eq \eqref{e110}
via the introduction of the $f$-cosmological constant and matter
sources as in Eq. \eqref{e114} is a solution of Eq. \eqref{e113}.
Hence, consequently the metrics $g$, and $f=\mathcal{C}^2g$ become
the solutions of the bigravity action Eq \eqref{e1}. Next, we will
construct an explicit example. Let us take the two proportional
metrics as FLRW type
\begin{subequations}\label{e115}
\begin{align}
g&=-dt^2+\frac{a^2(t)}{1-kr^2}dr^2+a^2(t)r^2 d\theta^2 +a^2(t)r^2
sin^2 \theta d\varphi^2,\notag\\
\notag\\
f&=-\mathcal{C}^2dt^2+\frac{\mathcal{C}^2a^2(t)}{1-kr^2}dr^2+\mathcal{C}^2a^2(t)r^2
d\theta^2 +\mathcal{C}^2a^2(t)r^2 sin^2 \theta d\varphi^2,
\tag{\ref{e115}}
\end{align}
\end{subequations}
with any of the proportionality constants $\mathcal{C}$
corresponding to the derived solutions $\{U_1,U_2,V_0,K,X_1\}$. We
also take the $g-$matter energy-momentum tensor in perfect fluid
form
\begin{equation}\label{e116}
T_M^g= \left(\begin{matrix}
-\rho(t)&\texttt{0}&\texttt{0}&\texttt{0}
\\
\texttt{0}&p(t)&\texttt{0}&\texttt{0}\\\texttt{0}&\texttt{0}&p(t)
&\texttt{0}\\\texttt{0}&\texttt{0}
&\texttt{0}&p(t)\end{matrix}\right),
\end{equation}
where we define the matrix $[T_M^g]^\mu_{\:\:\:\nu}\equiv
(T_M^g)^{\mu}_{\:\:\:\nu}$. Substituting $g$ from Eq.
\eqref{e115}, and also Eq. \eqref{e116} into Eq. \eqref{e112}
leads to the Friedmann equations
\begin{equation}\label{e117}
\big(\frac{\dot{a}}{a}\big)^2+\frac{k}{a^2}=\frac{8\pi
G}{3}\rho-\frac{\Lambda^g+m^2\tilde{\Lambda}}{6},
\end{equation}
and
\begin{equation}\label{e118}
\frac{2\ddot{a}}{a}=-\big(\frac{\dot{a}}{a}\big)^2-\frac{k}{a^2}-8\pi
Gp -\frac{\Lambda^g+m^2\tilde{\Lambda}}{2}.
\end{equation}
We also have the fluid equation
\begin{equation}\label{e119}
\dot{\rho}=-\frac{3\dot{a}}{a}\big(p+\rho\big),
\end{equation}
for the perfect fluid $g-$matter energy-momentum tensor in Eq.
\eqref{e116} coming from its conservation law $\nabla^\mu
T^{g}_{M\:\mu\nu}=0$. As usual, if one specifies the equation of
state $p=w\rho$ for the perfect fluid in Eq. \eqref{e116} one can
solve the scale factor $a(t)$ from the equations \eqref{e117},
\eqref{e118}, \eqref{e119},
 and then determine the metric $g$ in Eq. \eqref{e115} explicitly. If
 furthermore, via the fine-tuning that is introduced in Eq. \eqref{e114}
 we choose
\begin{equation}\label{e120}
T_M^f= \frac{\kappa}{\epsilon\,\mathcal{C}^2}\left(\begin{matrix}
-\rho(t)&\texttt{0}&\texttt{0}&\texttt{0}
\\
\texttt{0}&p(t)&\texttt{0}&\texttt{0}\\\texttt{0}&\texttt{0}&p(t)
&\texttt{0}\\\texttt{0}&\texttt{0}
&\texttt{0}&p(t)\end{matrix}\right),
\end{equation}
and $\Lambda^f=(\Lambda^g+m^2\tilde{\Lambda})/\mathcal{C}^2$ then
the $f$ metric in Eq. \eqref{e115} automatically satisfies Eq.
\eqref{e113} since now this equation becomes equivalent to Eq.
\eqref{e112} as we discussed above. In this manner, the metrics in
Eq. \eqref{e115} form explicit, and exact background solutions of
the bigravity action Eq. \eqref{e1} for any value of $\mathcal{C}$
coefficients of the cases $\{U_1,U_2,V_0,K,X_1\}$. We should
emphasize on an important point once more here, before we
conclude. Both in our example of cosmological solutions, and in
the more general proportional background cases given in Eq.
\eqref{e110}, the $g$, and $f$ metrics are defined in the same
coordinate chart. These solutions are similar to the first class
of solutions classified in \cite{8}. For this reason, we do not
have to solve a system of partial differential equations like the
cases in the second class of solutions listed in \cite{8} to find
the St\"{u}ckelberg fields which relate the $g$ metric components
to the $f$ metric ones which are diagonal in different coordinate
systems.
\section{Concluding Remarks}
Following the identification of a cubic matrix equation which
serves as the dynamical decoupling solution ansatz for the two
metrics of bigravity, we derived the general solutions of this
non-constant-coefficient matrix equation. We first obtained the
entire set of solutions of an ordinary cubic matrix equation by
classifying them into similarity equivalence classes of
diagonalizable and nondiagonalizable solutions. Later, starting
from these cases we have derived the complete set of roots of the
actual ansatz equation with coefficients also being functions of
the elementary symmetric polynomials of the roots. Consequently,
we obtained the Jordan canonical forms of all the equivalence
classes of solutions written in terms of the $\beta$-coupling
constants of the interaction term of the bigravity action. The
decoupling ansatz contributes an effective cosmological constant
to the foreground metric equation, while it leaves the background
metric theory as a detached GR. By being able to express the
elementary symmetric polynomials of the solutions of the cubic
matrix ansatz equation in terms of the coupling constants of the
theory we have also explicitly obtained the effective cosmological
constant for each class of solutions. Later on, we presented a
formal definition of the decoupling solution moduli of bigravity
which forms an important branch of exact solutions of the theory.
The analytical elements of this effective solution space can
explicitly be constructed following the solution of six algebraic
equations. In general, the decoupling solution space tells us
which couple of the metrics must be chosen to solve the
independent Einstein equations in two decoupled GR-sectors (one
having an always-nonvanishing cosmological constant due to the
presence of the effective contribution coming from the interaction
term). Finally, we presented a class of exact solutions in which
the two metrics are proportional to each other in the same
coordinate chart. We also explicitly constructed an example of the
cosmological proportional background solutions of the massive
bigravity. We should remark that the proportional background
solutions obtained in this work are new and they differ from the
already-existing ones in the literature \cite{8} which are derived
by excluding the possibility of the vanishing of the matrix $\tau$
which on the contrary sits at the focus of the present work.

In search for constructing self accelerating cosmologies, the
cosmological solutions of the ghost-free bigravity have attracted
a considerable attention in recent years
\cite{bac3,bm2,bm1,3,4,6,8,9,10,11,12}. The known class of
solutions studied in the corresponding literature can be divided
into three groups \cite{8}; the class in which both metrics are
proportional, a class of spherically symmetric solutions in which
the background metric is nondiagonal, and solutions including
diagonal but not proportional metrics. In this work, apart from
the parametrically-reduced cases, we construct the general
solution space of the bigravity field equations when the $g$, and
the $f$-sectors completely decouple from each other and simply
become GR with only an additional contribution of an effective
cosmological constant to the foreground metric Einstein equations.
Beside formally identifying the decoupling solution moduli of the
theory, we have presented a complete set of similarity equivalence
classes of analytic solutions up to solving a set of algebraic
equations. Each of these classes contains functionally infinite
degrees of freedom that generate proportional backgrounds,
diagonal, and nondiagonal solutions. We should also emphasize
that, clever choices of ansatz may substantially ease the general
explicit-solution building method we discussed. We believe that,
this complete classification of the decoupling solutions can be a
good starting point to study the viable cosmological solutions of
bigravity in a more systematical way.
\appendix
\section{Appendix}
In the Appendix, we present the collection of the matrices
constituting the set $\mathcal{J}$ defined in Eq. \eqref{e101}.
Their exact forms whose entries are functions of the
$\{\beta_i\}-$coefficients are derived in Section three by
substituting the corresponding ansatz into the equation
\eqref{e12}. Firstly,
\begin{equation}\label{a1}
 U_1=\left(\begin{matrix}
\lambda_{1}&\texttt{0}&\texttt{0}&\texttt{0}
\\
\texttt{0}&\lambda_{1}&\texttt{0}&\texttt{0}\\\texttt{0}&\texttt{0}&\lambda_{1}
&\texttt{0}\\\texttt{0}&\texttt{0}
&\texttt{0}&\lambda_{1}\end{matrix}\right),
U_2=\left(\begin{matrix}
\lambda_2&\texttt{0}&\texttt{0}&\texttt{0}
\\
\texttt{0}&\lambda_2&\texttt{0}&\texttt{0}\\\texttt{0}&\texttt{0}&\lambda_{2}
&\texttt{0}\\\texttt{0}&\texttt{0}
&\texttt{0}&\lambda_{2}\end{matrix}\right), U_3=
\left(\begin{matrix} \lambda_1&\texttt{0}&\texttt{0}&\texttt{0}
\\
\texttt{0}&\lambda_2&\texttt{0}&\texttt{0}\\\texttt{0}&\texttt{0}&\lambda_1
&\texttt{0}\\\texttt{0}&\texttt{0}
&\texttt{0}&\lambda_{2}\end{matrix}\right),
\end{equation}
where
\begin{subequations}\label{a2}
\begin{align}
\lambda_1&=\frac{\beta_2+\beta_3c_1-(\beta_3^2c_1^2+4\beta_2\beta_3c_1+\beta_2^2+4\beta_1\beta_3)^{1/2}}{2\beta_3},\notag\\
\lambda_2&=\frac{\beta_2+\beta_3c_1+(\beta_3^2c_1^2+4\beta_2\beta_3c_1+\beta_2^2+4\beta_1\beta_3)^{1/2}}{2\beta_3},
\tag{\ref{a2}}
\end{align}
\end{subequations}
with
\begin{equation}\label{a3}
c_1=\frac{-6\beta_2\beta_3\mp2(9\beta_2^2\beta_3^2-12\beta_1\beta_3^3)^{1/2}}{3\beta_3^2}.
\end{equation}
On the other hand,
\begin{equation}\label{a3.1}
 U_8=\left(\begin{matrix}
\lambda_{1}^{\prime}&\texttt{0}&\texttt{0}&\texttt{0}
\\
\texttt{0}&\lambda_{1}^{\prime}&\texttt{0}&\texttt{0}\\\texttt{0}&\texttt{0}&\lambda_{1}^{\prime}
&\texttt{0}\\\texttt{0}&\texttt{0}
&\texttt{0}&\texttt{0}\end{matrix}\right),\quad
U_{11}=\left(\begin{matrix}
\lambda_2^{\prime}&\texttt{0}&\texttt{0}&\texttt{0}
\\
\texttt{0}&\lambda_2^{\prime}&\texttt{0}&\texttt{0}\\\texttt{0}&\texttt{0}&\lambda_{2}^{\prime}
&\texttt{0}\\\texttt{0}&\texttt{0}
&\texttt{0}&\texttt{0}\end{matrix}\right),
\end{equation}
where
\begin{subequations}\label{a3.2}
\begin{align}
\lambda_1^{\prime}&=\frac{\beta_2+\beta_3c_2-((\beta_2+\beta_3c_2)^2+8\beta_1\beta_3+4\beta_2\beta_3c_2)^{1/2}}{2\beta_3},\notag\\
\lambda_2^{\prime}&=\frac{\beta_2+\beta_3c_2+((\beta_2+\beta_3c_2)^2+8\beta_1\beta_3+4\beta_2\beta_3c_2)^{1/2}}{2\beta_3},
\tag{\ref{a3.2}}
\end{align}
\end{subequations}
with
\begin{equation}\label{a3.3}
c_2=\frac{-3\beta_2\beta_3\pm
3(\beta_2^2\beta_3^2-\beta_1\beta_3^3)^{1/2}}{\beta_3^2}.
\end{equation}
Also,
\begin{equation}\label{a4}
 U=\left(\begin{matrix}
-\frac{\beta_2}{2\beta_3}&\texttt{0}&\texttt{0}&\texttt{0}
\\
\texttt{0}&-\frac{\beta_2}{2\beta_3}&\texttt{0}&\texttt{0}\\\texttt{0}&\texttt{0}&-\frac{\beta_2}{2\beta_3}
&\texttt{0}\\\texttt{0}&\texttt{0}
&\texttt{0}&\frac{3\beta_2+2\beta_3e_1}{2\beta_3}\end{matrix}\right),
\end{equation}
where $e_1$ is a free parameter such that $e_1\in
\big\{\mathcal{R}-\{-2\beta_2/\beta_3\}\big\}$.
\begin{equation}\label{a4.1}
 U^{\prime}=\left(\begin{matrix}
\texttt{0}&\texttt{0}&\texttt{0}&\texttt{0}
\\
\texttt{0}&-\frac{\beta_2}{\beta_3}&\texttt{0}&\texttt{0}\\\texttt{0}&\texttt{0}&-\frac{\beta_2}{\beta_3}
&\texttt{0}\\\texttt{0}&\texttt{0}
&\texttt{0}&\frac{2\beta_2+\beta_3e_1^{\prime}}{\beta_3}\end{matrix}\right),
\end{equation}
where $e_1^{\prime}$ is a free parameter such that
$e_1^{\prime}\in \big\{\mathcal{R}-\{-3\beta_2/\beta_3\}\big\}$.
In addition,
\begin{equation}\label{a4.2}
 U^{\prime\prime}=\left(\begin{matrix}
-\frac{\beta_1}{\beta_2}&\texttt{0}&\texttt{0}&\texttt{0}
\\
\texttt{0}&-\frac{\beta_1}{\beta_2}&\texttt{0}&\texttt{0}\\\texttt{0}&\texttt{0}&\texttt{0}
&\texttt{0}\\\texttt{0}&\texttt{0}
&\texttt{0}&\texttt{0}\end{matrix}\right).
\end{equation}
For the $V$-series we have,
\begin{subequations}\label{a5}
\begin{align}
 V_{0}&=\left(\begin{matrix}
-\frac{\beta_2}{2\beta_3}&\texttt{0}&\texttt{0}&\texttt{0}
\\
\texttt{0}&-\frac{\beta_2}{2\beta_3}&\texttt{0}&\texttt{0}\\\texttt{0}&\texttt{0}&-\frac{\beta_2}{2\beta_3}
&\texttt{0}\\\texttt{0}&\texttt{0}
&\texttt{0}&-\frac{\beta_2}{2\beta_3}\end{matrix}\right),V_{1}=\left(\begin{matrix}
-\frac{\beta_2}{2\beta_3}&\texttt{1}&\texttt{0}&\texttt{0}
\\
\texttt{0}&-\frac{\beta_2}{2\beta_3}&\texttt{0}&\texttt{0}\\\texttt{0}&\texttt{0}&-\frac{\beta_2}{2\beta_3}
&\texttt{0}\\\texttt{0}&\texttt{0}
&\texttt{0}&-\frac{\beta_2}{2\beta_3}\end{matrix}\right),
\notag\\
V_{2}&=\left(\begin{matrix}
-\frac{\beta_2}{2\beta_3}&\texttt{0}&\texttt{0}&\texttt{0}
\\
\texttt{0}&-\frac{\beta_2}{2\beta_3}&\texttt{1}&\texttt{0}\\\texttt{0}&\texttt{0}&-\frac{\beta_2}{2\beta_3}
&\texttt{0}\\\texttt{0}&\texttt{0}
&\texttt{0}&-\frac{\beta_2}{2\beta_3}\end{matrix}\right),V_{3}=
\left(\begin{matrix}
-\frac{\beta_2}{2\beta_3}&\texttt{0}&\texttt{0}&\texttt{0}
\\
\texttt{0}&-\frac{\beta_2}{2\beta_3}&\texttt{0}&\texttt{0}\\\texttt{0}&\texttt{0}&-\frac{\beta_2}{2\beta_3}
&\texttt{1}\\\texttt{0}&\texttt{0}
&\texttt{0}&-\frac{\beta_2}{2\beta_3}\end{matrix}\right),\notag\\
V_4&=\left(\begin{matrix}
-\frac{\beta_2}{2\beta_3}&\texttt{1}&\texttt{0}&\texttt{0}
\\
\texttt{0}&-\frac{\beta_2}{2\beta_3}&\texttt{0}&\texttt{0}\\\texttt{0}&\texttt{0}&-\frac{\beta_2}{2\beta_3}
&\texttt{1}\\\texttt{0}&\texttt{0}
&\texttt{0}&-\frac{\beta_2}{2\beta_3}\end{matrix}\right),
V_5=\left(\begin{matrix}
-\frac{\beta_2}{\beta_3}&\texttt{1}&\texttt{0}&\texttt{0}
\\
\texttt{0}&-\frac{\beta_2}{\beta_3}&\texttt{0}&\texttt{0}\\\texttt{0}&\texttt{0}&-\frac{\beta_2}{\beta_3}
&\texttt{0}\\\texttt{0}&\texttt{0}
&\texttt{0}&\texttt{0}\end{matrix}\right),\notag\\
V_6&=\left(\begin{matrix}
-\frac{\beta_2}{\beta_3}&\texttt{0}&\texttt{0}&\texttt{0}
\\
\texttt{0}&-\frac{\beta_2}{\beta_3}&\texttt{1}&\texttt{0}\\\texttt{0}&\texttt{0}&-\frac{\beta_2}{\beta_3}
&\texttt{0}\\\texttt{0}&\texttt{0}
&\texttt{0}&\texttt{0}\end{matrix}\right), \quad \quad \: \:
V_{10}= \left(\begin{matrix}
-\frac{\beta_2}{\beta_3}&\texttt{0}&\texttt{0}&\texttt{0}
\\
\texttt{0}&-\frac{\beta_2}{\beta_3}&\texttt{0}&\texttt{0}\\\texttt{0}&\texttt{0}&-\frac{\beta_2}{\beta_3}
&\texttt{0}\\\texttt{0}&\texttt{0}
&\texttt{0}&\texttt{0}\end{matrix}\right). \tag{\ref{a5}}
\end{align}
\end{subequations}
Now,
\begin{equation}\label{a6}
 Y=\left(\begin{matrix}
R&I&\texttt{0}&\texttt{0}
\\
-I&R&\texttt{0}&\texttt{0}\\\texttt{0}&\texttt{0}&R
&I\\\texttt{0}&\texttt{0} &-I&R\end{matrix}\right),
\end{equation}
where
\begin{equation}\label{a7}
R=-\frac{\beta_2}{2\beta_3},\quad
I=-\frac{(3\beta_2^2-4\beta_1\beta_3)^{1/2}}{2\beta_3}.
\end{equation}
The $X$-series elements in $\mathcal{J}$ are
\begin{subequations}\label{a8}
\begin{align}
 X_{1}&=\left(\begin{matrix}
-\frac{\beta_2}{3\beta_3}&\texttt{0}&\texttt{0}&\texttt{0}
\\
\texttt{0}&-\frac{\beta_2}{3\beta_3}&\texttt{0}&\texttt{0}\\\texttt{0}&\texttt{0}&-\frac{\beta_2}{3\beta_3}
&\texttt{0}\\\texttt{0}&\texttt{0}
&\texttt{0}&-\frac{\beta_2}{3\beta_3}\end{matrix}\right),X_{3}=\left(\begin{matrix}
-\frac{\beta_2}{\beta_3}&\texttt{0}&\texttt{0}&\texttt{0}
\\
\texttt{0}&-\frac{\beta_2}{\beta_3}&\texttt{0}&\texttt{0}\\\texttt{0}&\texttt{0}&\texttt{0}
&\texttt{0}\\\texttt{0}&\texttt{0}
&\texttt{0}&\texttt{0}\end{matrix}\right),
\notag\\
X_{4}&=\left(\begin{matrix}
-\frac{\beta_2}{2\beta_3}&\texttt{0}&\texttt{0}&\texttt{0}
\\
\texttt{0}&-\frac{\beta_2}{2\beta_3}&\texttt{0}&\texttt{0}\\\texttt{0}&\texttt{0}&-\frac{\beta_2}{2\beta_3}
&\texttt{0}\\\texttt{0}&\texttt{0}
&\texttt{0}&\texttt{0}\end{matrix}\right),\quad \: \: \: X_{7}=
\left(\begin{matrix}
-\frac{\beta_2}{\beta_3}&\texttt{0}&\texttt{0}&\texttt{0}
\\
\texttt{0}&-\frac{\beta_2}{\beta_3}&\texttt{0}&\texttt{0}\\\texttt{0}&\texttt{0}&\texttt{0}
&\texttt{1}\\\texttt{0}&\texttt{0}
&\texttt{0}&\texttt{0}\end{matrix}\right).\tag{\ref{a8}}
\end{align}
\end{subequations}
If we turn our attention on the $K$-series we have
\begin{equation}\label{a9}
 K=\left(\begin{matrix}
\tilde{\lambda}&\texttt{0}&\texttt{0}&\texttt{0}
\\
\texttt{0}&\tilde{\lambda}&\texttt{0}&\texttt{0}\\\texttt{0}&\texttt{0}&\tilde{\lambda}
&\texttt{0}\\\texttt{0}&\texttt{0}
&\texttt{0}&\tilde{\lambda}\end{matrix}\right),
\end{equation}
with
\begin{equation}\label{a10}
\tilde{\lambda}=-\frac{2}{7}\frac{\beta_2}{\beta_3}\pm\frac{1}{7}\bigg(4\frac{\beta_2^2}{\beta_3^2}-7\frac{\beta_1}{\beta_3}\bigg)^{1/2}.
\end{equation}
In addition, we also have
\begin{subequations}\label{a11}
\begin{align}
 K_{3}&=\left(\begin{matrix}
c_3&\texttt{0}&\texttt{0}&\texttt{0}
\\
\texttt{0}&c_3&\texttt{0}&\texttt{0}\\\texttt{0}&\texttt{0}&c_3
&\texttt{0}\\\texttt{0}&\texttt{0}
&\texttt{0}&-c_3\end{matrix}\right),K_{5}=\left(\begin{matrix}
-c_3&\texttt{0}&\texttt{0}&\texttt{0}
\\
\texttt{0}&c_3&\texttt{0}&\texttt{0}\\\texttt{0}&\texttt{0}&c_3
&\texttt{0}\\\texttt{0}&\texttt{0}
&\texttt{0}&c_3\end{matrix}\right),K^{\prime}=\left(\begin{matrix}
c_4&\texttt{0}&\texttt{0}&\texttt{0}
\\
\texttt{0}&c_4&\texttt{0}&\texttt{0}\\\texttt{0}&\texttt{0}&c_4
&\texttt{0}\\\texttt{0}&\texttt{0}
&\texttt{0}&\texttt{0}\end{matrix}\right),\notag\\K^{\prime\prime}&=\left(\begin{matrix}
c_5&\texttt{0}&\texttt{0}&\texttt{0}
\\
\texttt{0}&c_5&\texttt{0}&\texttt{0}\\\texttt{0}&\texttt{0}&\texttt{0}
&\texttt{0}\\\texttt{0}&\texttt{0}
&\texttt{0}&\texttt{0}\end{matrix}\right),\quad\:
K_{13}=\left(\begin{matrix} c_6&\texttt{0}&\texttt{0}&\texttt{0}
\\
\texttt{0}&c_6&\texttt{0}&\texttt{0}\\\texttt{0}&\texttt{0}&-c_6
&\texttt{0}\\\texttt{0}&\texttt{0}
&\texttt{0}&\texttt{0}\end{matrix}\right),\notag\\
K_{14}&=\left(\begin{matrix} c_7&\texttt{0}&\texttt{0}&\texttt{0}
\\
\texttt{0}&-c_7&\texttt{0}&\texttt{0}\\\texttt{0}&\texttt{0}&-c_7
&\texttt{0}\\\texttt{0}&\texttt{0}
&\texttt{0}&\texttt{0}\end{matrix}\right), \tag{\ref{a11}}
\end{align}
\end{subequations}
where
\begin{subequations}\label{a12}
\begin{align}
c_3&=-\frac{\beta_2}{\beta_3}\pm\bigg(\frac{\beta_2^2}{\beta_3^2}-\frac{\beta_1}{\beta_3}\bigg)^{1/2},\quad
c_4=-\frac{3}{8}\frac{\beta_2}{\beta_3}\pm\frac{1}{8}\bigg(9\frac{\beta_2^2}{\beta_3^2}-16\frac{\beta_1}{\beta_3}\bigg)^{1/2},\notag\\
c_5&=-\frac{1}{2}\frac{\beta_2}{\beta_3}\pm\frac{1}{2}\bigg(\frac{\beta_2^2}{\beta_3^2}-2\frac{\beta_1}{\beta_3}\bigg)^{1/2},\quad
c_6=-\frac{\beta_1}{\beta_2},\quad c_7=\frac{\beta_1}{\beta_2}.
\tag{\ref{a12}}
\end{align}
\end{subequations}
\section*{Acknowledgements}
We thank Merete Lillemark for useful communications.


\begin{thebibliography}{99}
\bibitem{BD1}
  Boulware D. G., and Deser S.
  \textit{``Can gravitation have a finite range?}'',
1972  \textit{Phys. Rev.} D {\textbf 6} 3368.
\bibitem{BD2}
  Boulware D. G., and Deser S.
  ``\textit{Inconsistency of finite range gravitation}'',
1972  \textit{Phys. Lett.} B {\textbf 40} 227.
\bibitem{dgrt1}
  de Rham C., and Gabadadze G.
  ``\textit{Generalization of the Fierz-Pauli Action}'',
   2010 \textit{Phys. Rev.} D {\textbf 82} 044020
  arXiv:1007.0443 [hep-th].
\bibitem{dgrt2}
  de Rham C., Gabadadze G., and Tolley A. J.
  ``\textit{Resummation of Massive Gravity}'',
  2011  \textit{Phys. Rev. Lett.}  {\textbf 106} 231101
  arXiv:1011.1232 [hep-th].
  \bibitem{fp}
  Fierz M., and Pauli W.
  \textit{``On relativistic wave equations for particles of arbitrary spin in an electromagnetic
  field}'',
1939   \textit{Proc. Roy. Soc. Lond.} A {\textbf173} 211.
\bibitem{hr1}
  Hassan S. F., and Rosen R. A.
  ``\textit{On Non-Linear Actions for Massive Gravity}'',
  2011 \textit{JHEP} {\textbf 1107} 009
  arXiv:1103.6055 [hep-th].
\bibitem{hr2}
  Hassan S. F., and Rosen R. A.
  ``\textit{Resolving the Ghost Problem in non-Linear Massive
  Gravity}'',
2012   \textit{Phys. Rev. Lett.}  {\textbf 108} 041101
  arXiv:1106.3344 [hep-th].
\bibitem{hr3}
  Hassan S. F., Rosen R. A., and Schmidt-May A.
  ``\textit{Ghost-free Massive Gravity with a General Reference
  Metric}'',
 2012  \textit{JHEP} {\textbf 1202} 026
  arXiv:1109.3230 [hep-th].
\bibitem{hrbg}
Hassan S. F., and Rosen R. A.
  ``\textit{Bimetric Gravity from Ghost-free Massive Gravity}'',
  2012 \textit{JHEP} {\textbf 1202} 126
  arXiv:1109.3515 [hep-th].
  \bibitem{bac1}
Baccetti V., Martin-Moruno P., and Visser M.
  ``\textit{Massive gravity from bimetric gravity}'',
 2013 \textit{Class. Quant. Grav.} {\textbf 30} 015004
arXiv:1205.2158 [gr-qc].
\bibitem{bac2}
Baccetti V., Martin-Moruno P., and Visser M.
  \textit{``Null Energy Condition violations in bimetric gravity''}, 2012
  \textit{JHEP} {\textbf 1208} 148
  arXiv:1206.3814 [gr-qc].
  \bibitem{bac3}
Baccetti V., Martin-Moruno P., and Visser M.
\textit{``Gordon and
Kerr-Schild ansatze in massive and bimetric
  gravity''}, 2012
  \textit{JHEP} {\textbf 1208} 108
  arXiv:1206.4720 [gr-qc].
\bibitem{bm2}
von Strauss M., Schmidt-May A., Enander J., Mortsell E., and
Hassan S. F. \textit{``Cosmological Solutions in Bimetric Gravity
and their Observational Tests''}, 2012 \textit{JCAP} \textbf{1203}
042 arXiv:1111.1655 [gr-qc].
\bibitem{bm1}
Volkov M. S. \textit{``Cosmological solutions with massive
gravitons in the bigravity theory''}, 2012 \textit{JHEP}
\textbf{1201} 035 arXiv:1110.6153 [hep-th].
\bibitem{massgrav}
  Y$\i$lmaz N. T.
  \textit{``Effective matter cosmologies of massive gravity I: non-physical
fluids''}, 2014 \textit{JCAP} \textbf{1408} 037
  arXiv:1405.6402 [hep-th].
  \bibitem{mgrphysfluid}
Y$\i$lmaz N. T.
  \textit{``Effective matter cosmologies of massive gravity: Physical
  fluids''}, 2014
  \textit{Phys. Rev.} {\textbf D90}, no.12 124034
arXiv:1412.4919 [hep-th].
\bibitem{3}
Volkov M. S. \textit{``Exact self-accelerating cosmologies in the
ghost-free bigravity and massive gravity''}, 2012 \textit{Phys.
Rev.} \textbf{D86} 061502 arXiv:1205.5713 [hep-th].
\bibitem{4}
Akrami Y., Koivisto T. S., and Sandstad M.
  \textit{``Accelerated expansion from ghost-free bigravity: a statistical analysis with improved
  generality''}, 2013
  \textit{JHEP} {\textbf 1303} 099
  arXiv:1209.0457 [astro-ph.CO].
\bibitem{6}
Volkov M. S.
  \textit{``Hairy black holes in the ghost-free bigravity theory''},
  2012
  \textit{Phys. Rev.} {\textbf D85} 124043
  arXiv:1202.6682 [hep-th].
  \bibitem{8}
Volkov M. S.
  \textit{``Self-accelerating cosmologies and hairy black holes in ghost-free bigravity and massive gravity''},
2013 \textit{Class. Quant. Grav.}  {\textbf 30} 184009
  arXiv:1304.0238 [hep-th].
\bibitem{9}
Koennig F., Patil A., and Amendola L. \textit{``Viable
cosmological solutions in massive bimetric gravity''}, 2014
  \textit{JCAP} {\textbf 1403} 029
  arXiv:1312.3208 [astro-ph.CO].
\bibitem{10}
De Felice A., G\"{u}mr\"{u}k\c{c}\"{u}o\u{g}lu A. E., Mukohyama
S., Tanahashi N., and Tanaka T.
  \textit{``Viable cosmology in bimetric theory''}, 2014
  \textit{JCAP} {\textbf 1406} 037
  arXiv:1404.0008 [hep-th].
\bibitem{11}
Koennig F., Akrami Y., Amendola L., Motta M., and Solomon A. R.
   \textit{``Stable and unstable cosmological models in bimetric massive gravity''}, 2014
  \textit{Phys. Rev.} {\textbf D90}, no.12 124014
  arXiv:1407.4331 [astro-ph.CO].
 \bibitem{12}
 Hassan S. F., Schmidt-May A., and von Strauss M.
  \textit{``Particular Solutions in Bimetric Theory and Their
  Implications''}, 2014
  \textit{Int. J. Mod. Phys.} {\textbf D23}, no.13 1443002
  arXiv:1407.2772 [hep-th].
\end{thebibliography}
\end{document}